%% file: root.tex
\documentclass[twocolumn]{article}
\usepackage[a4paper, total={7in, 10in}]{geometry}

\usepackage{cite}
\usepackage{amsmath,amssymb,amsfonts}
\usepackage{algorithmic}
\usepackage{graphicx}
\usepackage{textcomp}

\DeclareMathOperator*{\minimize}{\text{minimize}}
\DeclareMathOperator*{\subjto}{\text{subject to}}
\usepackage{physics}        
\usepackage{algorithm}      
\usepackage{multirow}       
\usepackage{url}
\usepackage{caption}
\usepackage{subcaption}
\usepackage{soulutf8}
\DeclareMathAlphabet{\mathpzc}{OT1}{pzc}{m}{it}

\newcommand{\todo}[1]{}
\usepackage{xcolor}
\definecolor{flame}{rgb}{0.89, 0.35, 0.13}

\newcommand{\changed}[1]{{\color{black} #1}}
\DeclareMathOperator*{\maximize}{maximize}
\newcommand{\dat}{\mathcal{D}}
\renewcommand{\todo}[1]{{\color{red} TODO: {#1}}}  
\newcommand{\state}{\mathpzc{x}}
\newtheorem{theorem}{Theorem}

\usepackage{abstract} 
\usepackage{titlesec} 
\renewcommand\thesection{\Roman{section}} 
\renewcommand\thesubsection{\Alph{subsection}} 
\titleformat{\section}[block]{\large\scshape\centering}{\thesection.}{1em}{} 
\titleformat{\subsection}[block]{\normalsize\it}{\thesubsection.}{1em}{} 
\usepackage[labelfont=bf]{caption}

\begin{document}
\title{Backflipping with Miniature Quadcopters by Gaussian Process Based Control and Planning}
\author{Péter Antal, Tamás Péni, 
and Roland Tóth
\thanks{The research was supported by the European Union within the framework of the National Laboratory for Autonomous Systems (RRF-2.3.1-21-2022-00002) and by the Eötvös Loránd Research Network (grant. number: SA-77/2021).}
\thanks{The authors are with the Systems and Control Laboratory, Institute for Computer Science and Control, Budapest, Hungary (email: antalpeter@sztaki.hu, peni@sztaki.hu, tothroland@sztaki.hu). R. T\'{o}th is also affiliated with the Vehicle Industry Research Center, Széchenyi István University and the Control Systems Group of the Eindhoven University of Technology.}}
\date{}

\maketitle

\begin{abstract}
The paper proposes two control methods~for performing a backflip maneuver with miniature quadcop\-ters. First, an existing feedforward control approach is improved {by finding the optimal sequence of motion primitives via Bayesian optimization, using a surrogate Gaussian Process model}. To evaluate the cost function, the flip maneuver is performed repeatedly in a simulation environment. The second method is based on closed-loop control and it consists of two main steps: first a novel robust, adaptive controller is designed to provide reliable reference tracking even in case of model uncertainties. The controller is constructed by augmenting the nominal model of the drone with a Gaussian Process that is trained by using measurement data. Second, an efficient trajectory planning algorithm is proposed, which designs feasible trajectories for the flip maneuver by using only quadratic programming. The two approaches are analyzed in simulations and in real experiments using Bitcraze Crazyflie 2.1 quadcopters.
\end{abstract}


\input{sec1-intro}

\input{sec2-GP}

\input{sec3-model}

\input{sec4-openloop}

\input{sec5-geometric}

\input{sec6-simu}

\input{sec7-experiments}

\input{sec8-summary}

{\footnotesize\bibliography{reference} }

\end{document}

%% file: sec1-intro.tex
\section{Introduction}
With the wide spread use of quadcopters, increasing expectations towards these systems  such as fast autonomous navigation in a cluttered environment, rapidly changing wind conditions in built environments, security and surveillance tasks,  
require to perform complex, 
fast maneuvers that push the drones to their physical limits \cite{wild}. In these cases, classical flight controllers designed for a linearized dynamical model of the vehicle are no longer sufficient and more advanced control methods capable to handle the entire operating domain are needed \cite{lelemc2010}. These algorithms can be developed based on nonlinear control techniques or machine learning approaches.

Execution of a backflip illustrates well such complex maneuvers because it requires careful handling of the full complex nonlinear behaviour of the drone and it is typically a challenging task even for a human pilot.
The complexity and speed of the maneuver is characterized by the fact that it takes less than a second to complete during which the vehicle is able to make a full turn around one of the horizontal axes.

Because of its benchmark characteristics, various control strategies has been already proposed to perform the flip maneuver. In \cite{energy-quaternion}, energy-based control is applied to overcome the uncontrollability of the quadcopter at singular configurations when following a circular or clothoidal reference trajectory. In \cite{lyapunov-flip}, a Lyapunov-stability based controller synthesis is used to execute multi-flip maneuvers with quadcopters. Machine learning approaches are utilized in many cases, for example to imitate the maneuver performed by an expert drone pilot with apprenticeship learning \cite{abbeel2010}, or train a deep neural network sensorimotor controller for executing acrobatic maneuvers \cite{deep_acrobatics}. 

A simple learning strategy for adaptive feedforward control is proposed in \cite{LSICRA2010}, based on the optimization of a parametric motion primitive sequence. As backflipping pushes the actuators of the quadcopter to their physical limits, the application of near-maximal and minimal control inputs are required. This approach builds on the theory of bang-bang control and first-principles motion primitive design to perform and optimize the flip maneuver. The 
method is easy to implement and it is well suited for generating a feasible motion sequence, however, many trials on the real robot are necessary to optimize the parameters of the motion and the resulting control law is sensitive to parameter uncertainties and external disturbances. These effects have greater influence on the behaviour of miniature quadcopters compared to  medium-sized and large drones, hence, the control robustness is even more important.

The robust adaptive control method we propose\footnote{In the conference paper \cite{antal2022}, preliminary version of the algorithms presented in this paper has been discussed. Compared to \cite{antal2022}, the main contributions of this paper include the introduction of the Bayesian optimization based feedforward method, the GP-based model augmentation, and the derivation and proof of the robustness properties of the feedback control algorithm.} in this paper is based on geometric control, which is a nonlinear approach for attitude feedback control of rigid bodies in 3D space. In \cite{lelemc2010}, it is theoretically proven that geometric control is able to stabilize the orientation of a quadcopter in the whole operating domain based on differential geometric considerations and Lyapunov stability. In \cite{Lee2013NonlinearRT}, {this} geometric control is augmented {with} robust terms to guarantee uniformly ultimately bounded tracking errors in the presence of {uncertainties} in the quadcopter dynamics. However, the control design requires a priori knowledge of the magnitude of external disturbances which can be challenging to forsee real-world situations. An adaptive augmentation of geometric control is proposed in \cite{goodarzi2015}, where the adaptive terms compensate the effects of uncertainties in the quadrotor dynamics, while the stability of the closed loop system is proven mathematically. Artificial neural networks are used in \cite{Bisheban2021} to develop an adaptive geometric control law that renders the quadcopter able to perform complex maneuvers in wind fields. Although both adaptive algorithms can be implemented efficiently, they do not provide an estimation of the uncertainty of the adaptive terms. Furthermore, both methods lack a systematic way to determine the parametric structure of the adaptive terms, which can be challenging without expert knowledge of the unmodeled dynamics and external disturbances. 
 
To overcome these challenges, the main contributions of our present work are as follows:
\begin{enumerate}
    \item[C1] \changed{We improve the convergence and the training time of the feedforward control 
    design of \cite{LSICRA2010} by applying Bayesian optimization with {a} \emph{Gaussian Process} (GP) surrogate function, making it possible to effectively apply the method in real experiments.}
    \item[C2] By {GP}-based augmentation of a nominal quadcopter model, we achieve adaption to unknown model dynamics and external disturbances together with  quantification of the remaining model uncertainty. A robust geometric control scheme is designed that exploits the GP model for improved robust performance compared to previous methods {and has convergence guarantees}.  
    \item[C3] To execute the flip maneuver by the robust geometric approach, we propose an optimization-based trajectory planning method. The algorithm is based on quadratic programming and it is computationally efficient.
    \item[C4] We compare the proposed methods in simulation and experimentally in performing the backflip maneuver. 
\end{enumerate}

The paper is structured as follows. First, a brief introduction to Gaussian Process regression is given in Section \ref{sec:GP}, while, in Section \ref{sec:model}, the dynamic motion model of quadcopters is presented. 
Then, the Bayesian optimization based improved feedforward control strategy is described in Section \ref{sec:flip}, corresponding to C1. The GP based robust geometric reference tracking control scheme and the quadratic programming based trajectory planning are presented in Section \ref{sec:geom}, corresponding to C2 and C3. 
Sections \ref{sec:simu} and \ref{sec:exp} give a detailed comparison of the two control approaches: first via simulations and then in real world experiments, providing C4. Finally in Section \ref{sec:conc}, conclusions on the proposed approaches are drawn.  A video presentation of our results is available at \url{https://youtu.be/Ed9jYlZr95c}. 


%% file: sec2-GP.tex
\section{Gaussian Process Regression}\label{sec:GP}
\emph{Gaussian Processes} (GPs) are universal function approximators  
\cite{RW2006}. Due to their flexibility, wide representation capability and ability to express uncertainty of the approximation, GPs have become popular in robotics and control engineering \cite{Zeilinger, Liu,Deisenroth2015, Torrente2021}. \changed{Other supervised learning approaches, such as artificial neural networks (ANNs) are frequently applied to compensate unmodeled dynamic effects (\cite{Bisheban2021,Bauersfeld2021}), however, the uncertainty quantification of ANNs is complex and unreliable. Basic statistical methods (e.g. Gamma tests) are not useful for robust control, parameter uncertainty with ellipsoidal regions is overly conservative, while both MCMC and dropout methods together with Bayesian ANNs are computationally overwhelming \cite{Abdar2021}. The main advantage of GPs over ANN based methods is that they provide co-estimation of the nominal functional relationship together with its uncertainty in a computationally efficient manner as shown in many applications \cite{RW2006}. This makes them especially attractive for developing adaptive robust control solutions. In terms of real-time implementation, the evaluation of baseline GP requires the entire training dataset that can be computationally demanding, however, there are several methods that solve this problem efficiently \cite{unifying_sparse}.}

GP regression is used for estimating an unknown, possibly nonlinear relation $f_0:\mathbb{R}^{n} \rightarrow \mathbb{R}$ between the input $x\in\mathbb{R}^{n}$ and noisy output observations $y\in\mathbb{R}$ of the form
\begin{align} \label{eq:datagen_relation}
    y = f_0(x) + \epsilon
\end{align}
where $\epsilon$ is an independent noise process with $\epsilon\sim \mathcal{N}(0, \sigma_\epsilon^2)$. In fact, $y$ and $\epsilon$ are random variables, but for the sake of simplicity, we will not use a different notation for their sample realization. 
Consider that a set of observations $\dat_N = \{ x_i,  y_i\}_{i=1}^N$ is available from \eqref{eq:datagen_relation}. 
The core idea of GP based estimation of  $f_0$ is to consider that candidate estimates $f$ belong to a GP, seen as a prior distribution.  
Then, using $\dat_N$ and this prior, a predictive GP distribution of $f$ is computed that provides estimate of $f_0$ in terms of its mean and describes uncertainty of this estimate by its variance.

In terms of definition, a \emph{Gaussian Process} $\mathcal{GP}: \mathbb{R}^d \rightarrow \mathbb{R}$ assigns to every point $x \in \mathbb{R}^n$ a random variable \mbox{$\mathcal{GP}(x) \in \mathbb{R}$} such that for any finite set $x_1 \ldots x_N$, the joint probability dis\-tri\-bution of $\mathcal{GP}\left(x_1\right), \ldots, \mathcal{GP}\left(x_N\right)$ is Gaussian. GPs are fully de\-termined by their mean $m$ and covariance functions $\kappa$, hence if $f\sim\mathcal{GP} (m, \kappa)$, 
\begin{align*} {m}(x) &=\mathbb{E}\{f(x)\} \\
{\kappa}(x,\tilde{x})& = 
\mathbb{E}\{(f(x)-m(x))(f(\tilde{x})-m(\tilde{x}))\},
\end{align*}
then the joint Gaussian probability of $\mathcal{GP}\left(x_1\right), \ldots, \mathcal{GP}\left(x_N\right)$ is $\mathcal{N}(M_x,K_{xx})$ with $M_x =\left[\ m\left(x_1\right)\ \cdots\ m\left(x_N\right)\ \right]^{\top}$ and $[K_{xx}]_{i,j} =\kappa\left(x_{i}, x_{j}\right)$, $i, j\in \{ 1\dots N\}$.
Both $m$ and $\kappa$, where the latter is also called a \emph{kernel} function, 
are often parametrized in terms of \emph{hyper parameters} $\theta\in\mathbb{R}^{n_\theta}$. In fact, taking $f\sim\mathcal{GP} (m, \kappa)$ as the prior distribution in the estimation process defines the prior knowledge about $f_0$ in terms of the mean function $m$, while the choice of $\kappa$ determines the function space in which an estimate of the function is searched for. Parametrization of $m$ and $\kappa$ in terms of $\theta$ allows to adjust the prior, i.e., these choices to the estimation problem of $f_0$ using $\dat_N$. For the estimation of a smooth $f_0$, a \emph{Squared Exponential} (SE) kernel for $\kappa$ is a common choice. The SE kernel is characterized by
\begin{align}\label{eq:sekernel}
    \kappa_\mathrm{SE}(x, \tilde{x}) = \sigma_\mathrm{f}^2 \exp \left( -\frac{1}{2}(x-\tilde{x})^\top \Lambda^{-1} (x-\tilde{x}) \right),
\end{align}
where hyper parameters are the scaling $\sigma_\mathrm{f}$, used for numerical conditioning, and the symmetric matrix $\Lambda$, which determines the smoothness of the candidate function class along each $x_i$.

Based on the given $\dat_N$ and the prior $f\sim\mathcal{GP} (m, \kappa)$, 
\begin{equation}\label{eq:prior}
   p(Y | X, \theta) = \mathcal{N}(M_x, K_{xx} + \sigma_\epsilon^2I),
\end{equation}
describes the probability density function of the outputs $Y=[\ y_1\ \cdots \ y_N\ ]^\top$ seen as random variables conditioned on the observed inputs $X=[\ x_1 \ \cdots \ x_N\ ]^\top$ and hyper parameter values $\theta$. 
To predict 
the value of the unknown function $f_0$ at a test point $x_*$, the following joint distribution 
\begin{align*}
   \begin{bmatrix}
 { Y} \\
 f(x_\ast) \\
 \end{bmatrix}
 \sim \mathcal{N}\left( \begin{bmatrix}
      M_x \\
      m(x_*)
   \end{bmatrix}, \begin{bmatrix}
     K_{xx} + \sigma_\epsilon^2I & K_x(x_*)  \\ K_x^\top (x_*)&   \kappa(x_*,x_*) 
   \end{bmatrix}
   \right)
\end{align*}
with $[ K_x(x_*) ]_i =\kappa(x_i,x_\ast)$ holds based on the previous considerations. Hence, the predictive distribution for $f(x_\ast)$, based on the observed samples $\{y_i\}_{i=1}^N$ in $
\dat_n$, is the posteriori $p (f(x_\ast) | \dat_N, x_\ast) = \mathcal{N}(\mu(x_*),\sigma(x_\ast))$ characterized by
\begin{subequations}\label{eq:posterior}
    \begin{align}
      \mu(x_*) &= m(x_\ast) +K_x^\top(x_\ast)\left(K_{xx}+\sigma^2_\epsilon I_N\right)^{\!-1}\!(Y\!-M_x) \label{eq:posterior_m} \\ 
       \sigma(x_*) &= 
      k(x_\ast,x_\ast)\!-\!{K_x^\top (x_\ast)}\left(K_{xx}+\sigma^2_\epsilon I_N\right)^{\!-1}\!K_x (x_\ast) \label{eq:posterior_s}
%
%
    \end{align}
\end{subequations}
The mean \eqref{eq:posterior_m} gives an approximation of $f_0(x_*)$ while the variance \eqref{eq:posterior_s} gives measure of the uncertainty of this approximation. Computation of \eqref{eq:posterior} requires only elementary matrix operations, therefore it is computationally efficient. 

To tune the hyper parameters $\theta$ and $\sigma_\epsilon^2$ associated with the prior, a common method is to maximize the likelihood, i.e. probability, of the observations of $Y$ for \eqref{eq:prior} marginalized w.r.t. (with respect to) $\theta$ and $\sigma_\epsilon$: 
\begin{align} \label{hyp:opt}
    \begin{bmatrix} \theta^* \\ \sigma_\epsilon^\ast \end{bmatrix} &= \arg \max_{\theta, \sigma_\epsilon} \ \log \left(p( Y | X, \theta, \sigma_\epsilon) \right),
\end{align}
where without loss of generality, the $\log$ of the pdf is taken to simplify the optimization problem and
\vspace{-2mm}
\begin{multline}
\log \left(p( Y | X, \theta,\sigma_\epsilon) \right)  = -\frac{1}{2} \left(Y^\top ( K_{xx}^{-1} + \sigma_\epsilon^2I_N) Y + \right. \\ \left. \log \det (  K_{xx}^{-1} + \sigma_\epsilon^2I_N) + N\log (2\pi) \right). 
\end{multline}
 For alternative methods, see \cite{RW2006}. Here, we considered scalar valued GPs, however, GP regression can {be} applied under multidimensional outputs by estimating a predictive distribution of each output dimension independently.  
 

%% file: sec3-model.tex
\section{Quadcopter Dynamics}\label{sec:model}
In this section, we introduce the basic principles of quadcopter modeling based on \cite{quad_model} and develop {a} nominal quadcopter model. 

\begin{figure}
    \centering 
    \includegraphics[width=.8\linewidth]{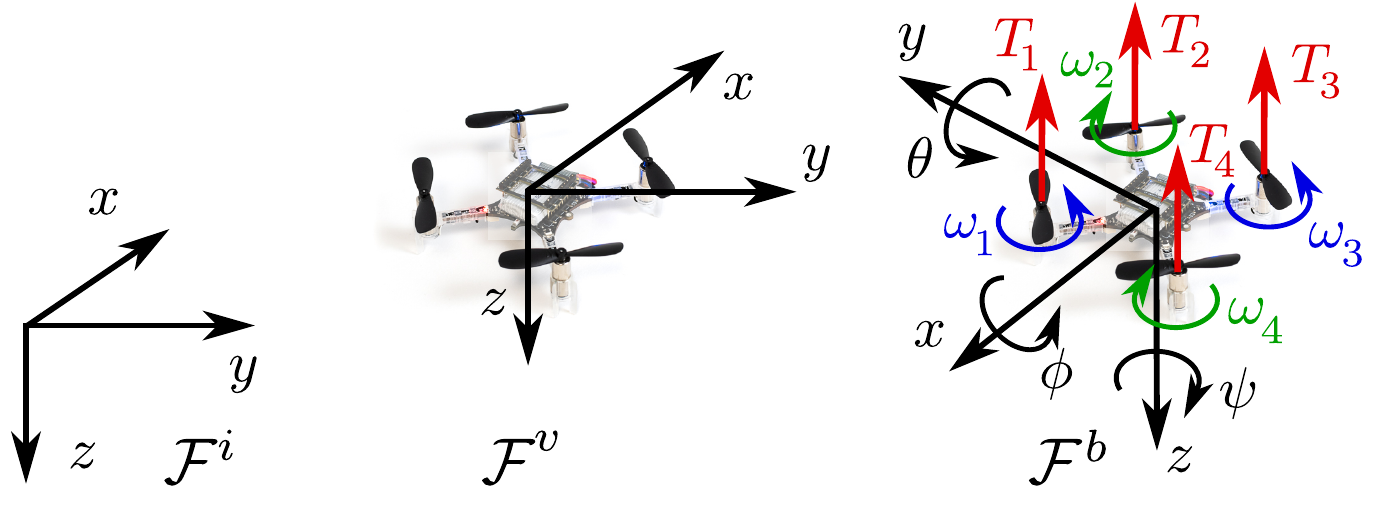}
    \caption[Inertial, vehicle, and body frames]{Inertial $\mathcal{F}^\mathrm{i}$, vehicle $\mathcal{F}^\mathrm{v}$, and body  $\mathcal{F}^\mathrm{b}$ frames describing the geometric relations of the vehicle and the environment. Thrusts and angular velocities of the rotors are also illustrated.}\label{fig:frames} \vspace{-4mm}
    \end{figure}

As the first step, three frames are introduced: the inertial frame $\mathcal{F}^\mathrm{i}$ in \emph{North-East-Down} (NED)  coordinates, the vehicle frame $\mathcal{F}^\mathrm{v}$ fixed to the vehicle, but aligned with $\mathcal{F}^\mathrm{i}$, and the body frame $\mathcal{F}^\mathrm{b}$ aligned with the body of the vehicle. The transformation from $\mathcal{F}^\mathrm{i}$ to $\mathcal{F}^\mathrm{v}$ is only a translation, while  $\mathcal{F}^\mathrm{v}$ and $\mathcal{F}^\mathrm{b}$ are connected by rotation only \cite{Beard2012}. In Fig.~\ref{fig:frames}, the three frames are displayed with the Euler angles characterizing the pose in the body frame (roll: $\phi$, pitch: $\theta$, yaw: $\psi$) together with the direction of the rotor thrusts and angular velocities. Based on these frames, the dynamic model 
is formulated as 
\begin{subequations}\label{eq:dyn}
    \begin{align}
        m\ddot{r} &= mge_3 - FR_\mathrm{b}^\mathrm{v} e_3,\label{eq:trans}\\
       \dot{R}_\mathrm{b}^\mathrm{v} & = R_\mathrm{b}^\mathrm{v}\hat{\omega}^\mathrm{b},\label{eq:rot1}\\
       J \dot{\omega}^\mathrm{b}  &= \tau - \omega^\mathrm{b}\times J^\mathrm{b} \omega^\mathrm{b},\label{eq:rot2}
      \end{align}
\end{subequations}
where $r=[\ x \ y \ z\ ]^\top$ is the position of the quadcopter in the inertial frame $\mathcal{F}^\mathrm{i}$, $e_3$ is the unit vector of the $z$ axis in $\mathcal{F}^\mathrm{b}$, $m$ is the mass of the drone, $F$ is the collective thrust of the propellers, and $g$ is the gravitational acceleration. $R_\mathrm{b}^\mathrm{v}(t)\in\mathrm{SO}(3)$ is the rotation matrix from $\mathcal{F}^\mathrm{b}$ to $\mathcal{F}^\mathrm{v}$, where SO(3) denotes the three-dimensional special orthogonal group, also called the \emph{rotation group}. 
Furthermore, $\omega^\mathrm{b}$ is the angular velocity of the vehicle in the body frame, $J^\mathrm{b}$ is the inertia matrix of the body of the vehicle, and $\tau=[\ \tau_\mathrm{x} \ \tau_\mathrm{y} \ \tau_\mathrm{z}\ ]^\top$ is the vector of torques produced by the propellers. The notation $\hat{\cdot}$ stands for the projection: $\mathbb{R}^3\rightarrow \mathrm{SO}(3)$ ensuring that $\hat{x}y = x\times y$ for all $x,y\in \mathbb{R}^3$ where $\times$ 
corresponds to the vector product.  
To simplify the notation, the coordinate frames are not indicated in the sequel, i.e. $R=R_\mathrm{b}^\mathrm{v},J=J^\mathrm{b}, \omega = \omega^\mathrm{b}$.

The dynamic model has four inputs, the collective thrust~$F$, and the torques around the three axes of the body frame $\tau$. \textcolor{black}{Assuming that the quadcopter configuration is symmetric and the torque generated by each propeller is proportional to the rotor thrust $T_i$, these inputs can be calculated as follows:} 
\begin{align}\label{eq:inputs}
    \begin{bmatrix}F \\ \tau\end{bmatrix}=\left[ \begin{array}{cccc}1 & 1 & 1 & 1 \\ -{l} &  -{l} &  {l} &  {l} \\  {l} & -{l} &  -{l} &  {l} \\ \frac{b}{c} & -\frac{b}{c} & \frac{b}{c} & -\frac{b}{c}\end{array}\right]\left[\begin{array}{c} T_1 \\ T_2 \\ T_3 \\T_4\end{array}\right],
    \end{align}
where $l$ is the distance of two motors along the $x$ axis, $b$ is the drag constant, and $c$ is the thrust constant. Furthermore, the thrust generated by each motor is considered to be proportional to the square of the corresponding angular velocity: $T_i = c\omega_i^2$ for $ i\in\{1,2,3,4\}$. In terms of input constraints, the individual rotor thrusts are in the range $0 \leq T_i \leq T_\mathrm{max}$, where $T_\mathrm{max}$ depends on the specific quadcopter design.

%% file: sec4-openloop.tex
\section{Feedforward Control by Bayesian Optimization}\label{sec:flip}
\subsection{Overview}
The flip maneuver can be executed as a 360 degree rotation around the $y$ axis of the body frame of the quadcopter, displayed in Fig. \ref{fig:frames}. If the corresponding ideal actuation sequence of the individual motors in terms of $T_{i}$ is computed based on the nominal quadcopter model to execute this maneuver, then the actuation sequence can be implemented on the real quadcopter in terms of feedforward control.   
However, computation of such an actuation profile is difficult, due to (i) the complexity of the involved optimization problem to find a feasible motion trajectory under the given actuation constraints and (ii) due to unmodeled aerodynamical effects that can significantly influence the system response.  

The feedforward control proposed in \cite{LSICRA2010} solves the problem above by tuning a fixed sequence of parameterized motion primitives via experiments. However, the approximation of the Jacobian matrix in the optimization loop requires many trials on the real drone and careful selection of the measurement data is needed to ensure the numerical convergence. We have modified the original algorithm 
at several points in order to improve its performance and adapt it to our specific design configuration. First, we tune the parameters of the motion primitives in simulation by using a high fidelity nonlinear model of the drone. Second, the optimization is solved by a Bayesian optimization method, using Gaussian Process surrogate function.
The main advantages of the proposed method are that Bayesian optimization is numerically better conditioned 
and requires significantly less function evaluations than Jacobi approximation as the evaluation points are systematically selected. This makes the proposed method computationally more favourable. 

\begin{figure}
\centering
\includegraphics[width=.2\textwidth]{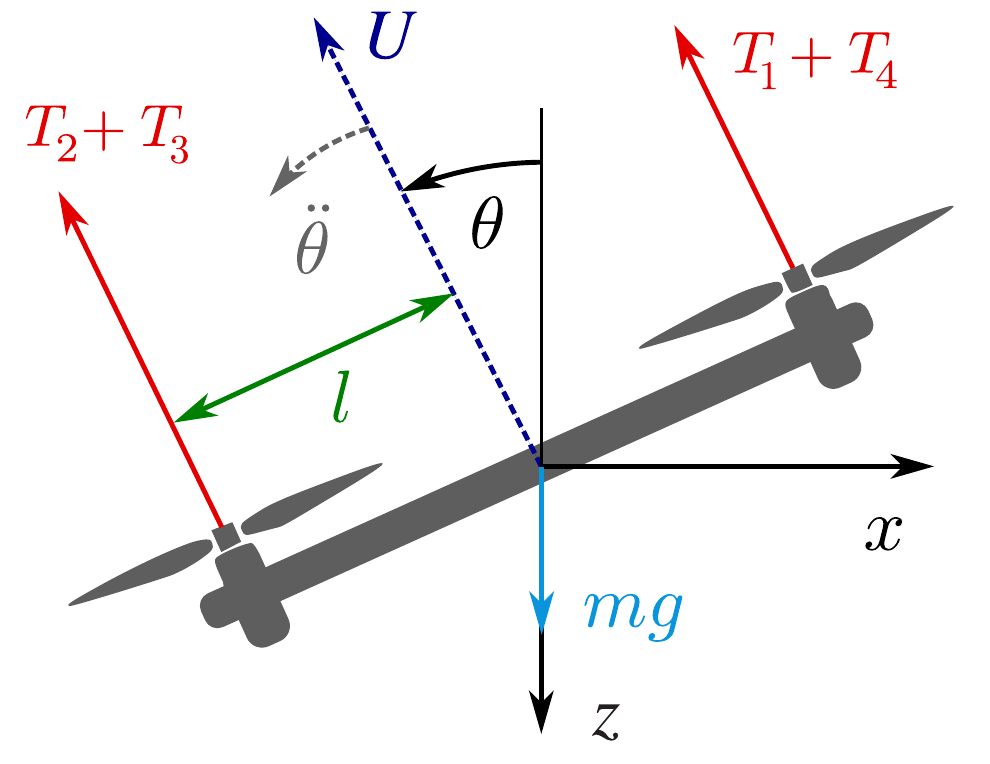}
\caption{The {2D} vehicle frame, the {pitch} orientation and {the} forces acting on the quadcopter {during an ideal flip}.}\label{fig:flipframe} \vspace{-4mm}
\end{figure}

Unlike in \cite{LSICRA2010}, we perform the backflip in a '$\times$' configuration as two rotors can produce a larger torque than one. The desired trajectory of the flip motion is within the $x-z$ plane of the body frame (illustrated in Fig.~\ref{fig:flipframe}), therefore the equations of motion \eqref{eq:dyn} can be simplified utilizing that the translation along the $y$ axis and rotation around the $x, z$ axes, i.e. the roll $\phi$ and the yaw $\psi$, are fixed to zero. 

The simplified equations of motion are as follows:
\begin{subequations}
  \begin{align}
m\ddot{x}&=-(T_1 + T_2 + T_3 + T_4)\sin\theta,\\
m\ddot{z}&=-(T_1 + T_2 +T_3 + T_4) \cos\theta+mg,\\
J_{\mathrm{yy}}\ddot{\theta} &= l(T_1+T_4-T_2-T_3).\label{eq:opinp1}
\end{align}  
\end{subequations}
The direction of the thrusts and pitch are illustrated in Fig.~\ref{fig:flipframe}. 

\begin{figure}
  \centering
  \includegraphics[width=.8\linewidth]{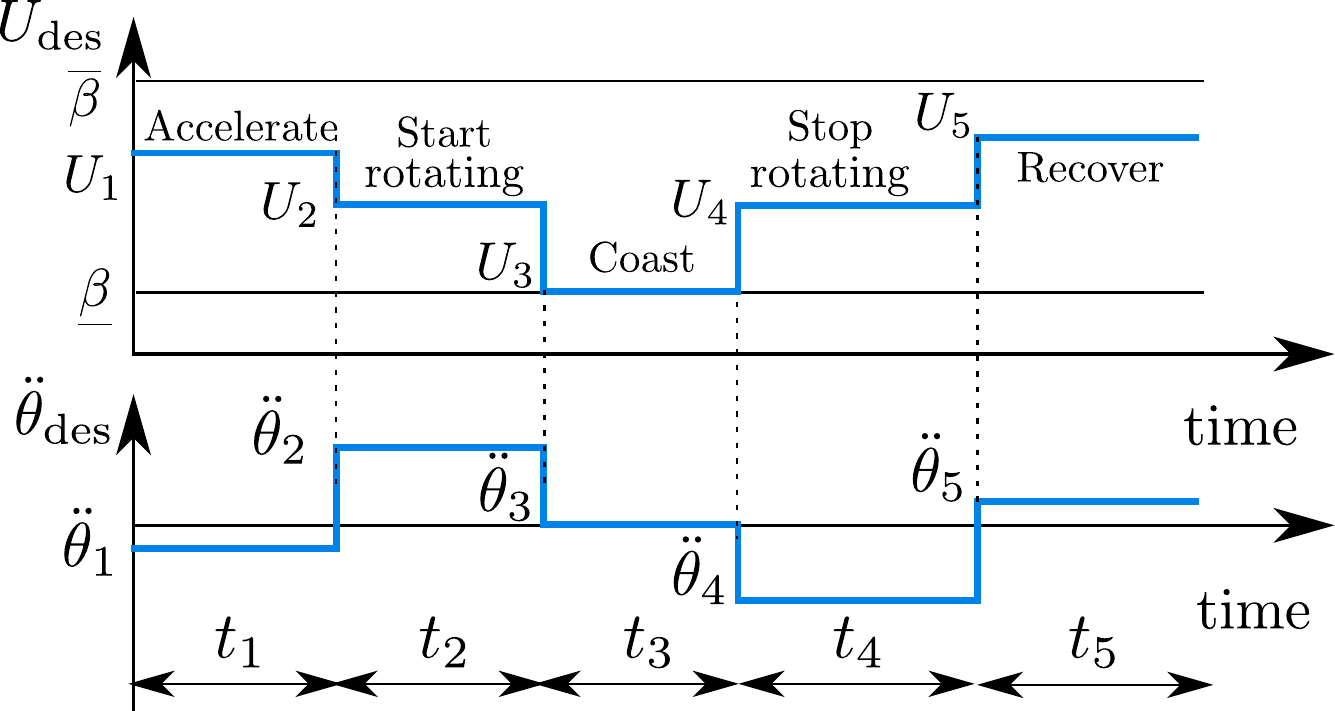}
  \caption{Flip motion described in terms of five parametrized motion primitives.}\label{fig:sections}\vspace{-4mm}
  \end{figure}

\subsection{Parametrized Primitives of the Maneuver}\label{sec:sections}
Similar to the control strategy laid down in \cite{LSICRA2010}, the backflip maneuver is divided to five main {phases (motion primitives)}  
that are illustrated in Fig. \ref{fig:sections} and defined as follows.
\changed{
\begin{enumerate}
\item \textbf{Accelerate:} Gain elevation and kinetic energy with near-maximal collective acceleration, while rotating slowly to the negative direction of the pitch angle $\theta$.
\item \textbf{Start Rotation:} Increase angular velocity by applying maximal differential thrust.
\item \textbf{Coast:} With low and uniform thrusts, hold the angular velocity, wait for the drone to rotate.
\item \textbf{Stop Rotation:} Use maximal differential thrust to decrease angular velocity and stop the rotation.
\item \textbf{Recover:} Apply near-maximal collective thrust to compensate gravity, and try to get back to hover mode.
\end{enumerate}}

Each of the five {phases} 
has three parameters, the collective acceleration $U_i$, duration $t_i$, and angular acceleration $\ddot{\theta}_i$, resulting in 15 parameters altogether. However, based on \cite{LSICRA2010}, the number of parameters can be reduced by applying bang-bang type control on a restricted control envelope. 
This means that the control actions $U_i$ and $\ddot\theta_i$ are either zero or near-maximal {during all phases}. 
As a result, only the following 5 independent parameters remain: 
$\eta=\begin{bmatrix} U_1 & t_1 & t_3 & U_5& t_5
\end{bmatrix} ^\top \in \mathbb{R}_+^5$.

These parameters are tuned to minimize the norm of the final state error $e\in\mathbb{R}^5$, which is obtained by applying the actuation profile in Fig.~\ref{fig:sections} in open loop and taking the difference between the final and the initial state. 
Formally, the optimization problem can be written as follows:
\begin{align}
\minimize_{\eta \in   \mathbb{R}_+^5} \quad& \| e(\eta) \|_2\label{eq:optim}  \\
\text{s.t.} \quad&e(\eta) =  [\ x(t_\mathrm{f}) \ z(t_\mathrm{f}) \ \dot{x}(t_\mathrm{f}) \ \dot{z}(t_\mathrm{f}) \ \theta(t_\mathrm{f})\ ]^\top, \nonumber\\
&x(t_0)=z(t_0)=\dot{x}(t_0)=\dot{z}(t_0)=0,\nonumber\\
&U_\mathrm{min}\leq U_i\leq U_\mathrm{max}\quad\; i\in\{1,5\},\nonumber\\
&t_\mathrm{min}\leq t_j \leq t_\mathrm{max}\quad\quad j\in\{1,3,5\},\nonumber
\end{align} 
where $t_0$ and $t_\mathrm{f}$ are the initial and final time instants of the maneuver and {the} bounds
$U_\mathrm{min},\; U_\mathrm{max},\; t_\mathrm{min}$, $t_\mathrm{max}$ are determined from the physical limitations of the drone.

\subsection{Bayesian Parameter Optimization}
In \cite{LSICRA2010}, the numerical optimization of $\eta$ is based on iterative optimization, using an approximate Jacobian matrix of the final state error w.r.t. the parameter vector. However, the numerical gradient approximation has vast computational cost, because the whole maneuver needs to be simulated in every approximation step and it suffers from convergence problems. Here, as our first contribution, we apply a Bayesian optimization approach to find the global optimum of \eqref{eq:optim}. This approach does not require the calculation of derivatives and it is suitable for global optimization of cost functions that are expensive to evaluate, see \cite{frazier2018tutorial, brochu2010tutorial}. To apply this approach, the optimization problem \eqref{eq:optim} is written as 
\begin{align}
 \maximize_{x\in \mathcal{X}} f(x),
\end{align}
where $\eta = x \in\mathbb{R}^n$ is the $n=5$ dimensional vector of optimization variables, $\mathcal{X}$ is the feasible parameter set (bounded interval of the search space for the parameters $\eta$), and $f$ is the objective function, i.e., $f(x)=-\|e(\eta=x)\|_2$. 
The core concept of Bayesian optimization is to evaluate the unknown objective function at limited number of points, giving $\dat_N=\{y_i=f(x_i),x_i\}_{i=1}^N$, fit a surrogate model based on GP regression on the data, and optimize  this surrogate model of the original objective function \cite{RW2006}. The optimization is performed iteratively, where in each step the next evaluation point is determined by minimizing a so called \emph{acquisition function}, the objective function is evaluated at this point and the surrogate model is updated. The optimization stops if the minimum is reached with high confidence or the iteration reaches a certain number of evaluations. 

The acquisition function blends the approximated objective (the mean of the GP) and the approximation uncertainty (the variance of the GP) in a scalar valued function that can be optimized by standard gradient-based procedure. 
A common acquisition function is \textit{expected improvement}, defined as follows. After $f$ is evaluated in $N$ points, giving the observation data set $\dat_N$, a GP predictive distribution  
$\hat{f}_N \sim \mathcal{GP}(\mu_{N},\sigma_{N})$ is obtained w.r.t. $\dat_{N}$. Let $\bar{f}_N^+$ be the value of the best sample so far and $x_N^+ = \arg \max_{x\in X_N} \mu^{(N)}(x)$
be the location of that sample based on $X_N=\{x_i\}_{i=1}^N$, i.e., $\bar{f}_N^+=\mu^{(N)}(x_N^+)$.  
The next test point $x_{N+1}$ is chosen such that the expected improvement predicted by the GP model is the best w.r.t. $x^+_N$:
\begin{equation}\label{eq:ei}
    \mathrm{EI}_N(x):=\mathbb{E}\left\{\lfloor \hat{f}_N(x)-\bar{f}_N^+\rfloor \right\},
\end{equation}
where $\lfloor y \rfloor = \max(y,0) $. The right hand side of \eqref{eq:ei} has an analytical form and the next test point is obtained via $x_{N+1} = \arg \max_{x\in\mathcal{X}} \mathrm{EI}_N(x),$ the point with highest expected improvement. In the literature, there are other common acquisition functions, e.g. upper confidence bound or knowledge gradient 
\cite{frazier2018tutorial}. Steps of {the} Bayesian optimization using a GP surrogate model {are summarized} in Alg.~\ref{alg:bayes}, based on \cite{frazier2018tutorial}.

\begin{algorithm}[!t]
\caption{High-level steps of Bayesian optimization}
\label{alg:bayes}
\begin{algorithmic}
    \STATE $f$ is evaluted at $N>0$ initial points, providing $\dat_N$
    \STATE Set a GP prior on $f$ in terms of $\hat{f}\sim\mathcal{GP} (m, \kappa)$ 
    \WHILE {$N\leq N_\mathrm{max}$}
    \STATE Determine $\hat{f}_N \sim \mathcal{GP}(\mu_{N},\sigma_{N})$ via \eqref{eq:posterior} and \eqref{hyp:opt} w.r.t. $\dat_{N}$ 
    \STATE Let $x_{N+1} = \arg \max_{x\in\mathcal{X}} \mathrm{EI}_N(x),$ 
    \STATE Observe $y_{N+1} = f(x_{N+1})$ 
    \STATE $\dat_{N+1}=\dat_N \cup \{y_{N+1},x_{N+1}\}$
    \STATE $ N \leftarrow N+1$
	\ENDWHILE
	\RETURN  $x_{\ast} = \arg \max_{x\in\mathcal{X}} \mu^{(N)}(x)$
\end{algorithmic}
\end{algorithm}



With the mathematical model of the quadcopter and a suitable optimization algorithm, it is possible to simulate the maneuver with different parameter sets, optimize the motion, and implement it on the vehicle. For the implementation of the flip maneuver, a stabilizing feedback controller is also required to balance the quadcopter at the beginning and after the end of the maneuver, for which we use the geometric control method introduced in Section~\ref{sec:geom}.


%% file: sec5-geometric.tex
\section{Geometric Tracking Control with Trajectory Planning}\label{sec:geom}

The second considered approach for quadcopter backflipping is based on closed-loop control. For this purpose, as our second contribution, we propose a novel robust adaptive controller to provide reliable reference tracking even in the case of modeling uncertainties and external disturbances. The proposed method is an extension of the geometric control law \cite{Lee2013NonlinearRT} applicable for trajectory tracking of aggressive maneuvers. Furthermore, we also introduce a novel optimization-based trajectory planning method for this geometric approach, which is essential for finding an efficient motion path for backflipping. 



\subsection{Robust Adaptive Geometric Control by Gaussian Processes}
The proposed robust nonlinear geometric tracking control is based on the extension of the control law introduced in \cite{Lee2013NonlinearRT}. The control method is able to track a reference position $r_\mathrm{d}(t) =[\ x_\mathrm{d}(t)\ y_\mathrm{d}(t)\ z_\mathrm{d}(t)\ ]^\top $ and a reference attitude $R_\mathrm{d}(t)\in \mathrm{SO}(3)$, represented by rotation matrices. To synthesize the control law, we use \eqref{eq:dyn}, describing the quadcopter dynamics, and augment it by an additive state-dependent disturbance: 
\begin{subequations}\label{eq:uncdyna}
    \begin{align}
        m\ddot{r} &= mge_3 - FRe_3 + \Delta_\mathrm{r}(\state ),\label{eq:gpsys1}\\
        J\dot{\omega} &= \tau - \omega \times J\omega  + \Delta_\mathrm{R}(\state ),\label{eq:gpsys2}
      \end{align}
\end{subequations}
\changed{where $\Delta_\mathrm{r}(\state ), \Delta_\mathrm{R}(\state ) \in \mathbb{R}^3$ comprise the model errors and uncertainties associated with the quadcopter dynamics, which depend on the state vector $\state=[\ r^\top\; \dot r^\top\; q^\top\; \omega^\top\ ]^\top$. The attitude quaternion $q$ is computed directly from $R$.}

For controlling the flight dynamics in \eqref{eq:uncdyna}, we propose to use the control law in \cite{Lee2013NonlinearRT} augmented by the adaptive state-dependent terms $\eta_\mathrm{r}, \eta_\mathrm{R}$ to cope with $\Delta_\mathrm{r}$ and $\Delta_\mathrm{R}$:
\begin{subequations}\label{eq:geomlaw}
    \begin{align}
        F =&  (-k_\mathrm{r}e_\mathrm{r} - k_\mathrm{v}e_\mathrm{v} - mge_3 + m\ddot{r}_\mathrm{d} - \eta_\mathrm{r} + \mu_\mathrm{r})^\top R e_3,\\
        \begin{split}\label{eq:geomtau}
            \tau =&  -k_\mathrm{R} e_\mathrm{R} - k_\Omega e_\Omega + \omega \times J\omega -\\& J\left(\hat{\Omega} R^\top\! R_\mathrm{d} \Omega_{\mathrm{d}}-R^\top R_{\mathrm{d}} \dot{\Omega}_{\mathrm{d}}\right) - \eta_\mathrm{R} + \mu_\mathrm{R},
        \end{split}      
      \end{align}
\end{subequations}
where $k_\mathrm{r}, k_\mathrm{v}, k_\mathrm{R}, k_\omega \in \mathbb{R}$ are the controller gains and 
\begin{subequations}\label{eq:geomerrors}
\begin{align}
    e_\mathrm{r} &= r-r_\mathrm{d}, & e_\mathrm{R} &= \frac{1}{2}\left(R_\mathrm{d}^\top R - R^\top R_\mathrm{d}\right)^\vee,\\
    e_\mathrm{v} &= \dot r - \dot r_\mathrm{d}, &  e_\omega &= \omega - R^\top R_\mathrm{d}\omega_\mathrm{d},
\end{align}
\end{subequations}
are error terms with $r_\mathrm{d}$, $R_\mathrm{d}$ and $\omega_\mathrm{d}$ corresponding to the position, orientation and angular velocity references, $\mathrm{tr}(\cdot)$ is the trace operator, and the \emph{vee operator} $(\cdot)^\vee:\mathrm{SO}(3)\rightarrow \mathbb{R}^3$ is the inverse of the hat operator $\hat{(\cdot)}$. The attitude tracking error $e_\mathrm{R}$ is interpreted as the gradient of the attitude error function characterized by \cite{lelemc2010}
\begin{align}\label{eq:psi}
    \Psi(R, R_\mathrm{d}) = \frac{1}{2}\mathrm{tr}\left( I-R_\mathrm{d}^\top \mathrm{R}\right).
\end{align}
Furthermore, the angular velocity error term satisfies the equation $\dot\Psi = e_\mathrm{R}^\top e_\omega$.

We identify the external disturbances from noisy observations using Gaussian Processes, more specifically in the form 
\begin{subequations}
   \begin{align}
    &\hat\Delta_\mathrm{r} = \mathcal{GP}_\mathrm{r}(\state ) \sim \mathcal{N}(\eta_\mathrm{r}(\state ), \Sigma_\mathrm{r}(\state )),\\
    &\hat\Delta_\mathrm{R} = \mathcal{GP}_\mathrm{R}(\state ) \sim \mathcal{N}(\eta_\mathrm{R}(\state ), \Sigma_\mathrm{R}(\state )).
\end{align} 
\end{subequations}
The mean of the GP-s is then directly used in \eqref{eq:geomlaw} to compensate the effect of {$\Delta_\mathrm{r}(\state ), \Delta_\mathrm{R}(\state )$}. The uncertainty of the approximations, characterized by the covariances $\Sigma_\mathrm{r}$, $\Sigma_\mathrm{R}$ is handled by introducing the additional terms $\mu_\mathrm{r}$ and $\mu_\mathrm{R}$ that make the controller robust to this uncertainty. To define  these terms, we assume that the {GPs} are trained until the true 
$\Delta_\mathrm{r}(\state ), \Delta_\mathrm{R}(\state )$ are inside the 95\% confidence interval. We can now define $\mu_\mathrm{r}$ and $\mu_\mathrm{R}$ similarly to \cite{Lee2013NonlinearRT}, as
    \begin{align}
        \mu_{\mathrm{r}} &=-\frac{\delta_{\mathrm{r}}^{\tau+2} e_{\mathrm{B}}\left\|e_{\mathrm{B}}\right\|^{\tau}}{\delta_{\mathrm{r}}^{\tau+1}\left\|e_{\mathrm{B}}\right\|^{\tau+1}+\epsilon_{\mathrm{r}}^{\tau+1}}, &
        e_{\mathrm{B}} &=e_{\mathrm{v}}+\frac{c_{1}}{m} e_{\mathrm{r}}, \label{eq:robust_terms}\\
        \mu_{\mathrm{R}} &=-\frac{\delta_{\mathrm{R}}^{2} e_{\mathrm{A}}}{\delta_{\mathrm{R}}\left\|e_{\mathrm{A}}\right\|+\epsilon_{\mathrm{R}}}, &
        e_{\mathrm{A}} &=e_{\Omega}+c_{2} J^{-1} e_{\mathrm{R}},\nonumber
    \end{align}
where $ c_{1}, c_{2}, \epsilon_{\mathrm{r}}, \epsilon_{\mathrm{R}}, \tau$ are positive constants, $\tau>2$, $\delta_\mathrm{r}, \delta_\mathrm{R}$ are the uncertainty bounds and $\lVert\cdot\rVert$ is the Euclidean vector norm. However, instead of estimating $\delta_\mathrm{r}, \delta_\mathrm{R}$ as in \cite{Lee2013NonlinearRT}, we utilize the uncertainty of the corresponding trained Gaussian Process. We calculate the standard deviation of a GP at an evaluation point {as} the {norm of the} square root of the covariance matrix {(sphere bounding the ellipsoidal level set):}
\begin{subequations}
    \begin{align}
    \sigma_\mathrm{r}(\state ) &= \norm{L_\mathrm{r}(\state )}_2 = \sqrt{\lVert \Sigma_\mathrm{r}(\state )\rVert_2}, \quad L_\mathrm{r}L_\mathrm{r}^\top=\Sigma_\mathrm{r},\\
    \sigma_\mathrm{R}(\state ) &= \norm{L_\mathrm{R}(\state )}_2 = \sqrt{\lVert\Sigma_\mathrm{R}(\state )\rVert_2}, \quad L_\mathrm{R} L_\mathrm{R}^\top=\Sigma_\mathrm{R},
\end{align}
\end{subequations}
where $\lVert\cdot\rVert_2$ is the 
2,2 induced norm (also called spectral norm) of a matrix which is equal to its largest singular value. We define the uncertainty bounds using the 95\% confidence interval of the normal distribution, namely 
\begin{align}
    \hat\delta_\mathrm{r}(\state ) = 2\sigma_\mathrm{r}(\state ),\quad \hat\delta_\mathrm{R}(\state ) = 2\sigma_\mathrm{R}(\state ).
\end{align}
From the confidence interval, we calculate an ultimate bound for the difference between the disturbances and the adaptive terms over the operating domain as follows:
\begin{align} \label{eq:ua:bound}
    \delta_\mathrm{r} = \max_\state \hat\delta_\mathrm{r}(\state ),\quad \delta_\mathrm{R} = \max_\state \hat\delta_\mathrm{R}(\state ).
\end{align}
Now we can show for the considered uncertainty bound \eqref{eq:ua:bound} that the following stability guarantee holds true:


\begin{theorem} Consider that the control force $F$ and torque $\tau$ defined by \eqref{eq:geomlaw} are applied on the uncertain system \eqref{eq:uncdyna}. Given any $\psi_{1,\max}, e_\mathrm{r_{max}} > 0$ and initial conditions that satisfy \vspace{-4mm}
\begin{subequations}\label{eq:bounds}\begin{align}
    \Psi(R(0), R_\mathrm{d}(0)) & < \psi_{1,\max} < 1,\\
    \norm{e_\mathrm{r}(0)} & < e_\mathrm{r_{max}},
\end{align}\end{subequations}
{then} there exists a controller (in terms of the choice of parameters $k_\mathrm{r}$, $k_\mathrm{v}$, $k_\mathrm{R}$, $k_\omega$, $c_{1}$, $c_{2}$, $\epsilon_{\mathrm{r}}$, $\epsilon_{\mathrm{R}}$, $\tau$) such that all error terms in \eqref{eq:geomerrors} are uniformly ultimately bounded.
\end{theorem}

\textit{Proof}: 
Throughout the proof, we adapt the steps of Proposition 3 in \cite{Lee2013NonlinearRT} to the GP based uncertainty bounds. 
First, we derive the dynamics of the rotational and translational tracking error and construct Lyapunov functions for them. 

Based on \eqref{eq:uncdyna}, \eqref{eq:geomlaw}, \eqref{eq:geomerrors}, the error dynamics can be expressed in the following form:
\begin{subequations}
\begin{align}
      m\dot{e}_\mathrm{v} &= mge_3-m\ddot{r}_\mathrm{d} - \frac{F}{e_3^\top R_\mathrm{d}^\top R e_3}R_\mathrm{d}e_3 - X  + \Delta_\mathrm{r}\nonumber\\
 &=-k_\mathrm{r} e_\mathrm{r} - k_\mathrm{v} e_\mathrm{v} - X + \Delta_\mathrm{r} - \eta_\mathrm{r} + \mu_\mathrm{r},\\
    X &= \frac{F}{e_3^\top R_\mathrm{d}^\top R e_3}((e_3^\top R_\mathrm{d}^\top R e_3)Re_3-R_\mathrm{d} e_3),\label{eq:proof_x}\\
 J\dot{e}_\omega & =  \tau + \Delta_R - \omega \times J\omega +  J\left(\hat{\omega} R^\top R_{\mathrm{d}} \omega_{\mathrm{d}}-R^\top R_{\mathrm{d}} \dot{\omega}_{\mathrm{d}}\right) \nonumber\\&= -k_\mathrm{R}e_\mathrm{R}-k_\omega e_\omega + \Delta_\mathrm{R} - \eta_\mathrm{R} + \mu_\mathrm{R}.
\end{align}
\end{subequations}
Similar to \cite{Lee2013NonlinearRT}, consider the following Lyapunov function candidates:
\begin{subequations}
  \begin{align}
    \mathcal{V}_1 &= \frac{1}{2}k_\mathrm{r}\norm{e_\mathrm{r}}^2 + \frac{1}{2}m\norm{e_\mathrm{v}}^2 + c_1e_\mathrm{r}^\top e_\mathrm{v},\label{eq:lyapcan1}\\
    \mathcal{V}_2 &= \frac{1}{2}e_\omega^\top J e_\omega + k_\mathrm{R}\Psi(R, R_\mathrm{d}) + c_2 e_\mathrm{R}^\top e_\omega.\label{eq:lyapcan2}
\end{align}  
\end{subequations}
First, let us focus on $\mathcal V_2$. For the attitude error function $\Psi$, a lower and upper bound can be given in terms of the attitude error $e_\mathrm{R}$ as follows:
\begin{align}\label{eq:psibounds}
    \frac{1}{2}\left\|e_{\mathrm{R}}\right\|^{2} \leq \Psi\left(R, R_{\mathrm{d}}\right) \leq \frac{1}{2-\psi_{1,\max}}\left\|e_{\mathrm{R}}\right\|^{2},
\end{align}
The detailed derivation of these bounds can be found in \cite{lee2011}. Note that, \eqref{eq:psibounds} implies that $\Psi$ is positive definite and decrescent for all $t\in\mathbb{R}_+$. By using \eqref{eq:psibounds}, the following upper bound can be derived for the time derivative of $\mathcal{V}_2$:
\begin{align}\label{eq:lyap1}
    &\dot{\mathcal{V}}_2 \leq -z_2^\top W_2 z_2 + e_\mathrm{A}^\top (\Delta_\mathrm{R} - \eta_\mathrm{R} + \mu_\mathrm{R}),\\
    z_2 = & \begin{bmatrix} \| e_\mathrm{R} \| \\ \| e_\omega \| \end{bmatrix}\in (\mathbb{R}^2)^{\mathbb{R}_+},\quad W_{2}=\left[\begin{array}{cc}
\frac{c_{2} k_\mathrm{R}}{\lambda_\mathrm{M}} & -\frac{c_{2} k_{\omega}}{2 \lambda_\mathrm{m}} \\
-\frac{c_{2} k_{\omega}}{2 \lambda_\mathrm{m}} & k_{\omega}-c_{2}
\end{array}\right],\nonumber
\end{align}
where $\lambda_\mathrm{m}, \lambda_\mathrm{M}$ denote the smallest and largest eigenvalue of the inertia matrix $J$. If the controller parameters are chosen such that $W_2$ is positive definite, then $z_2^\top W_2 z_2$ is a positive definite function. Now we examine the second term on the r.h.s of \eqref{eq:lyap1} and construct an upper bound for this term as well. 
First, note that $\norm{\Delta_\mathrm{R} - \eta_\mathrm{R}}\leq \delta_\mathrm{R}$ holds 
under the assumption that $\Delta_\mathrm{r}, \Delta_\mathrm{R}$ are inside the 95\% confidence interval of the GP distribution. 
By using this inequality and the definition of the robust control law  \eqref{eq:robust_terms}, the following upper bound can be obtained:
\begin{align}\label{eq:bound1}
    \begin{split}
        e_\mathrm{A} ^\top  (\Delta_\mathrm{R} - \eta_\mathrm{R} + \mu_\mathrm{R}) &\leq \delta_\mathrm{R} \| e_\mathrm{A} \| - \frac{\delta_\mathrm{R}^2 \| e_\mathrm{A} \|^2}{\delta_\mathrm{R} \| e_\mathrm{A} \| + \epsilon_\mathrm{R}} \\
        &= \frac{\delta_\mathrm{R} \| e_\mathrm{A} \|}{\delta_\mathrm{R} \| e_\mathrm{A} \| + \epsilon_\mathrm{R}} \epsilon_\mathrm{R} \leq \epsilon_\mathrm{R},
    \end{split}
\end{align}
\changed{where $\epsilon_\mathrm{R}$ is chosen to be a sufficiently small positive constant, therefore}
\begin{align}\label{eq:v2dot}
    \dot{\mathcal{V}}_2 \leq -z_2^\top W_2 z_2 + \epsilon_\mathrm{R}.
\end{align}
\changed{The right hand side of \eqref{eq:v2dot} is a shifted negative definite function, hence the tracking errors $e_\mathrm{R}$ and $e_\omega$ are uniformly ultimately bounded.}

Consider now Lyapunov function candidate $\mathcal{V}_1$.  Its time derivative can be given as follows:   
\begin{align}\label{eq:v1dot}
\begin{split}
   \dot{\mathcal{V}}_{1}=&-\left(k_\mathrm{v}-c_{1}\right)\left\|e_\mathrm{v}\right\|^{2}-\frac{c_{1} k_\mathrm{r}}{m}\left\|e_\mathrm{r}\right\|^{2}-\frac{c_{1} k_\mathrm{v}}{m} e_\mathrm{r} ^\top e_\mathrm{v} \\
&+\left(X+\Delta_\mathrm{r}-\eta_\mathrm{r}+\mu_\mathrm{r}\right)^\top \left(\frac{c_{1}}{m} e_\mathrm{r}+e_\mathrm{v}\right).
\end{split}
\end{align}
Similarly to the case of $\mathcal{V}_2$, the first three terms can be made negative definite by suitably choosing the parameters of the controller.  The last term can be divided into two parts: the first one is $e_\mathrm{B}^\top (\Delta_\mathrm{r} - \eta_\mathrm{r} + \mu_\mathrm{r})$ and $e_\mathrm{B}^\top X$ is the second. By using \eqref{eq:ua:bound} and \eqref{eq:robust_terms}, an upper bound can be derived for the first term:
\begin{align}\label{eq:epsilon_r}
    \begin{split}
        e_\mathrm{B}^\top (\Delta_\mathrm{r} - \eta_\mathrm{r} + \mu_\mathrm{r}) &  \leq \delta_\mathrm{r} \| e_\mathrm{B} \| - \frac{\delta_\mathrm{r}^{\tau+2} \| e_\mathrm{B} \|^{\tau+2}}{\delta_\mathrm{r}^{\tau+1} \| e_\mathrm{B} \|^{\tau+1} + \epsilon_\mathrm{r}^{\tau+1}} \\
        &= \frac{\delta_\mathrm{r} \| e_\mathrm{B} \|}{\delta_\mathrm{r}^{\tau+1} \| e_\mathrm{B} \|^{\tau+1} + \epsilon_\mathrm{r}^{\tau+1}} \epsilon_\mathrm{r}^{\tau+1} \leq \epsilon_\mathrm{r},
    \end{split}
\end{align}
\changed{where similarly to $\epsilon_\mathrm{R}$, $\epsilon_\mathrm{r}$ is also a sufficiently small positive constant, both typically in the order of magnitude of $10^{-4}$ to $10^{-2}$.} To obtain an upper bound for $e_\mathrm{B}^\top X$, first we construct an upper bound for $X$ by using \eqref{eq:proof_x} : 
\begin{align}\label{eq:xbound}
 & \| X \| \leq \|A \| \|(e_3^\top R_\mathrm{d}^\top R e_3)Re_3-R_\mathrm{d} e_3 \|\\ &\leq (k_\mathrm{r}\norm{e_\mathrm{r}} + k_\mathrm{v}\norm{e_\mathrm{v}} + B + \delta_\mathrm{r})\norm{(e_3^\top R_\mathrm{d}^\top R e_3)Re_3-R_\mathrm{d} e_3}, \nonumber
\end{align}
where $A =  -k_\mathrm{r} e_\mathrm{r} - k_\mathrm{v} e_\mathrm{v} - mge_3 +m\ddot{r}_\mathrm{d} - \eta_\mathrm{r} + \mu_\mathrm{r}$ and we assume that the reference trajectory $r_\mathrm{d}$ has been designed such that condition $\| -mge_3 + m\ddot{r}_\mathrm{d} -\eta_\mathrm{r} \| < B$ holds for some $B>0$. By using the following relation (see \cite{Lee2013NonlinearRT} for details),
\begin{align}
\begin{split}
   &\left\|\left(e_{3}^\top R_\mathrm{d}^\top  R e_{3}\right) R e_{3}-R_\mathrm{d} e_{3}\right\|  \leq\left\|e_\mathrm{R}\right\|=\sqrt{\Psi(2-\Psi)} \\
& \leq\left\{\sqrt{\psi_{1, \mathrm{max}}\left(2-\psi_{1, \mathrm{max}}\right)} \triangleq \alpha\right\}<1,
\end{split}
\end{align}
the upper bound for $X$ can be expressed as follows: 
\begin{align}\label{eq:alpha}
    \| X \| \leq (k_\mathrm{r}\norm{e_\mathrm{r}} + k_\mathrm{v}\norm{e_\mathrm{v}} + B + \delta_\mathrm{r})\alpha.
\end{align}
By substituting \eqref{eq:epsilon_r} and \eqref{eq:alpha} into \eqref{eq:v1dot}, we obtain
\begin{align}\label{eq:v1dot2}
   \dot{\mathcal{V}}_{1} \leq &-\left(k_\mathrm{v}(1-\alpha)-c_{1}\right)\left\|e_\mathrm{v}\right\|^{2}-\frac{c_{1} k_\mathrm{r}}{m}(1-\alpha)\left\|e_\mathrm{r}\right\|^{2} \nonumber \\
&+\left\|e_\mathrm{R}\right\|\left\{\left(B+\delta_\mathrm{r}\right)\left(\frac{c_{1}}{m}\left\|e_\mathrm{r}\right\|+\left\|e_\mathrm{v}\right\|\right)+k_\mathrm{r}\left\|e_\mathrm{r}\right\|\left\|e_\mathrm{v}\right\|\right\} \nonumber \\
&+\frac{c_{1} k_\mathrm{v}}{m}(1+\alpha)\left\|e_\mathrm{r}\right\|\left\|e_\mathrm{v}\right\| + \epsilon_\mathrm{r}. 
\end{align}
According to the proof of Proposition 3 in \cite{Lee2013NonlinearRT}, inequality \eqref{eq:v1dot2} implies that the tracking errors $e_\mathrm{r}$, $e_\mathrm{v}$ are uniformly ultimately bounded as well. This completes the proof. 
\hfill $\square$

\changed{\textit{Remark:} The user-{specified tolerable error bounds} $\psi_{1,\max}, e_\mathrm{r_{max}}$ in \eqref{eq:bounds} are required to be realistic in order to achieve high control performance {in the allowed input range}.}

\subsection{Trajectory Planning for the Flip Maneuver}
In order to use the geometric tracking controller to perform the flip maneuver, a suitable reference trajectory is needed. For this purpose, we introduce an optimization based trajectory design method which is an additional contribution of the paper.

Based on the controller structure \eqref{eq:geomlaw}, first an attitude reference trajectory $R_\mathrm{d}$ is constructed and then it is completed with a position reference $r_\mathrm{d}$. 
Similarly to the feedforward approach, the objective for trajectory planning is that the quadcopter should arrive as close to the starting point as possible, while keeping the control inputs within the allowed range during the maneuver. 

\begin{figure}
    \centering
    \includegraphics[width=.8\linewidth]{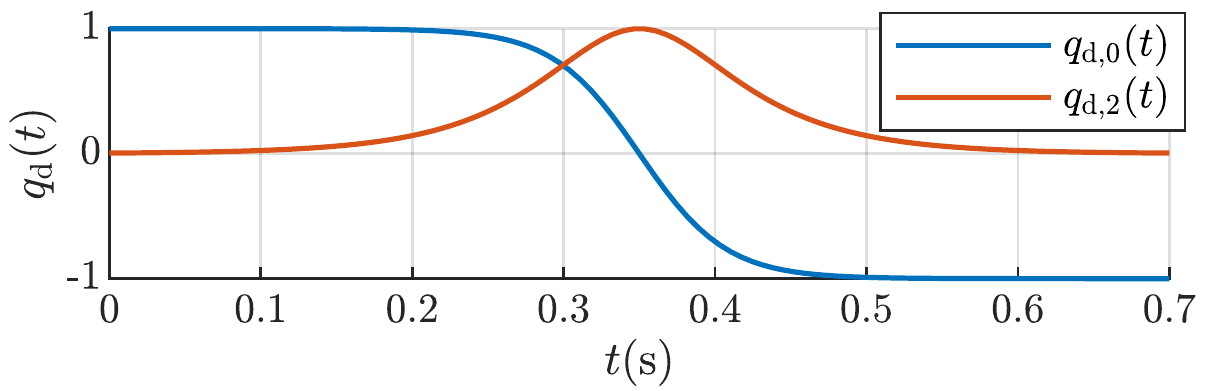}
    \caption{Attitude quaternion reference trajectory for the backflip maneuver with $\nu_\mathrm{m}=35$ 1/s, $t_\mathrm{m} = 0.7$ s.}
    \label{fig:attref} \vspace{-3mm}
\end{figure}

The attitude reference is specified in unit quaternions: 
$q_\mathrm{d} = [\ q_{\mathrm{d},0}\ q_{\mathrm{d},1}\ q_{\mathrm{d},2}\ q_{\mathrm{d},3}\ ]^\top$, where $q_{\mathrm{d},0}$ is the scalar part of the quaternion, and $q_{\mathrm{d},2}$ corresponds to the pitch angle, as $q_{\mathrm{d},1}=q_{\mathrm{d},3}=0$, because both the roll and yaw angles are zero during the flip. Utilizing that $q_\mathrm{d}$ is a unit quaternion, we can express the third element of it as $q_{\mathrm{d},2} = \sqrt{1-q_{\mathrm{d},0}^2}$, 
hence it is sufficient to design a trajectory only for $q_{\mathrm{d},0}$. 
 A 360 degree rotation around the $y$ axis means that the scalar part of the attitude quaternion goes from 1 to -1. In the trajectory design it is important to stay within the $q_{\mathrm{d},0}(t)\in [-1, 1]$ range, because only unit quaternions describe rotation. For this purpose, the smooth sigmoid function \vspace{-2mm}
\begin{equation}\label{eq:sigmoid}
    q_{\mathrm{d},0}(t) = \frac{2}{1+e^{-\nu_\mathrm{m}\left(t-\frac{t_\mathrm{m}}{2}\right)}}-1
\end{equation}
is chosen to describe the scalar part of the reference attitude, where the parameters are the horizontal scaling of the sigmoid curve $\nu_\mathrm{m}$ and the execution time $t_\mathrm{m}$. The attitude quaternion reference trajectory is displayed in Fig. \ref{fig:attref}. Assuming that $\phi\equiv\psi\equiv 0$ during the flip, the conversion to Euler angles yields $\theta=2\mathrm{arccos}(q_{0,\mathrm{d}})$, where $\theta(t)\in [-\pi, \pi]$. Hence the pitch angle goes smoothly from zero to $\pi$, jumps to $-\pi$, and goes smoothly to zero.
Besides of rotation, the maneuver also requires translational motion, because without proper lifting at the beginning, the quadcopter would fall to the ground due to gravity. The position reference is designed considering that the rotational and translational equations of the dynamical model are coupled. The translational motion of the flip maneuver is within the $x-z$ plane, therefore $y_\mathrm{d}(t) = 0$. The other two equations of the translational dynamics in \eqref{eq:trans} are
\begin{subequations}\label{eq:any}
    \begin{align}
       m \ddot x & = - FR_{1,3},\\
        m \ddot{z} & = - F R_{3,3} + mg,  
    \end{align}
\end{subequations}
where $R_{i,j}$ denotes the $(i, j)$-th entry of the rotation matrix~$R$. However, assuming that the attitude {rapidly} converges 
to the reference, we can substitute the reference rotation matrix in \eqref{eq:any}, resulting in the translational state-space representation
\begin{align}\label{eq:linsys}
    &\dot{\zeta} = A\zeta + B u,\\
    &\zeta=\begin{bmatrix} x \\ \dot x \\ \tilde z \\ \dot{\tilde{z}}\end{bmatrix},\quad A = \begin{bmatrix} 0 & 1 & 0 & 0 \\ 0 & 0 & 0 & 0 \\ 0 & 0 & 0 & 1 \\ 0 & 0 & 0 & 0 \end{bmatrix},\quad B =\frac{1}{m}\begin{bmatrix} 0 \\ R_{\mathrm{d},1,3} \\ 0 \\ R_{\mathrm{d},3,3}\end{bmatrix},\nonumber
\end{align}
where $\zeta$ is the reduced state vector capturing the translational motion, $R_{\mathrm{d},i,j}$ are the corresponding elements of the reference rotation matrix $R_\mathrm{d}$ (converted from the reference quaternion $q_\mathrm{d}$). 
As the motion equations are decoupled, the effect of gravity can be added to $z$ 
which is denoted by $\tilde{z}$ (modified state) in the equation. Notice that \eqref{eq:linsys} is a \emph{Linear Time-Varying} (LTV) state-space representation with the thrust force $F=u$ as the only control input. 

By discretizing \eqref{eq:linsys} using complete, zero-order hold discretization with discretisation step size $T_\mathrm{s}>0$:
\begin{equation}
    \begin{split}
    &\zeta_{k+1} = A_\mathrm{d} \zeta_k + B_{\mathrm{d},k} u_k,\\
    &A_\mathrm{d}= \begin{bmatrix} 1 & T_\mathrm{s} & 0 & 0 \\ 0 & 1 & 0 & 0 \\ 0 & 0 & 1 & T_\mathrm{s} \\ 0 & 0 & 0 & 1 \end{bmatrix}\!,  \ \ B_{\mathrm{d},k}=\frac{T_\mathrm{s}}{m}\arraycolsep=1.4pt\def\arraystretch{1.3}\left[\begin{array}{c} \tfrac{T_\mathrm{s}}{2} R_{\mathrm{d},1,3}(k) \\ R_{\mathrm{d},1,3}(k) \\ \tfrac{T_\mathrm{s}}{2}R_{\mathrm{d},3,3}(k) \\R_{\mathrm{d},3,3}(k) \end{array}\right]\!,
\end{split}
\end{equation}
where $k\in\mathbb{Z}$ denotes the discrete time, i.e. $\zeta_k$ expresses $\zeta(kT_\mathrm{s})$,
a quadratic programming problem can be formulated over a finite horizon, similarly to model predictive control, to find a motion trajectory for executing the flip. The input of the model is the collective thrust of the propellers, $u_k=F_k$. 
 For a fixed duration of the maneuver with $N$ discrete time steps, the following quadratic optimization problem is formulated:
    \begin{align} 
        \minimize_{\{u_k\}_{k=1}^N} \;\; &\sum_{k=1}^N \left[ \left(\zeta_k-\zeta_{\mathrm{d},k}\right)^{\!\top}\!  Q_k  \left(\zeta_k-\zeta_{\mathrm{d},k}\right) + u_k^\top W_k u_k\right]\nonumber\\
        \subjto \;\; & \zeta_{k+1} =  A_\mathrm{d} \zeta_k + B_{\mathrm{d},k} u_k,\label{eq:quadprog}\\
        &\{ \zeta_{k}\}_{k=1}^N \in \mathcal{X},\quad \{u_{k}\}_{k=0}^N \in \mathcal{U},\nonumber
    \end{align}
where $ Q_k\in \mathbb{R}^{4\times 4}$ and $W_k \in\mathbb{R}$ are weight matrices, $\zeta_0$ is the initial state, and $\mathcal{X}, \mathcal{U}$ are constraint sets for the states and the control input, respectively. The only objective of the trajectory design is to minimize the final position error of the quadcopter and keep the position within a specified range, therefore the weight matrices are $(W_k ,Q_k)= (0,0)$ for $k=1\dots N-1$, except for the weight of the final state that is $ Q_N = \mathrm{diag}(1, 0, 1, 0)$ while $W_N=0$. As all the other weights are zero, it is only required to define a final state position reference $\zeta_{\mathrm{d},N}$, the components of which are zero except for the effect of the gravity in $\tilde{z}_{\mathrm{d},N} = 0.5T_\mathrm{s}^2 N^2 g$. 

We specify linear constraints for the states: $x\in[x_-,x_+],$ $z\in[z_-, z_+]$ to model the available space for the maneuver, {preventing}  collisions with other objects or walls. We also define linear constraints for the control input, namely
    \begin{align}
        \frac{\norm{\tau_k}}{l} \leq u_k = F_k \leq F_\mathrm{max}-\frac{\norm{\tau_k}}{l},
    \end{align}
where $\tau_k$ is the vector of the three torques around the three body axes, out of which $\tau_{\mathrm{x},k}=\tau_{\mathrm{z},k}=0$ normally during the flip, $l$ is the distance of the quadcopter center of mass and the propellers projected to the $x-z$ plane, and $F_{\mathrm{max}}$ is the maximal collective thrust of the rotors. The torque control input $\tau_k$ is calculated from the reference attitude $R_\mathrm{d}$ based on \eqref{eq:geomtau} assuming that the rotation errors $e_\mathrm{R},e_\omega$ are zero.

The optimization problem in \eqref{eq:quadprog} can be solved easily by using an off-the-shelf QP solver, e.g. by \verb+quadprog+ in Matlab. Finally, to get a smooth trajectory, we fit cubic splines on the discrete reference points $\{\zeta_k\}_{k=0}^N$ obtained in \eqref{eq:quadprog}. 

%% file: sec6-simu.tex
\section{Simulation Study}\label{sec:simu}

\subsection{Environment}
To analyze the properties and the performance of the introduced methods, we tested them in a simulation environment based on the dynamic model of a Bitcraze Crazyflie 2.1 miniature quadcopter. The same drone is used in real experiments, presented in Section~\ref{sec:exp}. For both simulation and control design, the considered physical parameters of the quadcopter are given in Table~\ref{tab:params}, which are based on \cite{Forster}. 
\begin{table}[t]
    \centering
    \setlength{\tabcolsep}{1.5pt}
    \caption{Considered physical parameters of the Crazyflie~2.1.}
    \label{tab:params}
    \begin{tabular}{|l|c|rl|}
        \hline
        \phantom{o}Mass &\phantom{o} $m$ \phantom{o}& 28&g \\
        \hline
        \phantom{o}Prop-to-prop length \phantom{o}&\phantom{o} $l$\phantom{o} & 92&mm\\
        \hline
        \multirow{3}{*}{\phantom{o}Diagonal inertia} & \phantom{o}$J_{\mathrm{xx}}$\phantom{o} & $1.4\cdot 10^{-5}$&kgm$^2$\\
         \cline{2-4}
        &\phantom{o} $J_{\mathrm{yy}}$ \phantom{o}& $1.4\cdot 10^{-5}$&kgm$^2$\\
        \cline{2-4}
        & \phantom{o}$J_{\mathrm{zz}}$\phantom{o} & \phantom{o}$2.17\cdot 10^{-5}$&kgm$^2$\phantom{o}\\
        \hline
        \phantom{o}Thrust coefficient & $c$ & $2.88\cdot 10^{-8}$ & Ns$^2$\\
        \hline
        \phantom{o}Drag coefficient & $b$ & \phantom{o}$7.24\cdot 10^{-10}$ & Nms$^2$\phantom{o}\\
        \hline
    \end{tabular} 
\end{table}

\begin{figure}
\centering
\includegraphics[scale=.48]{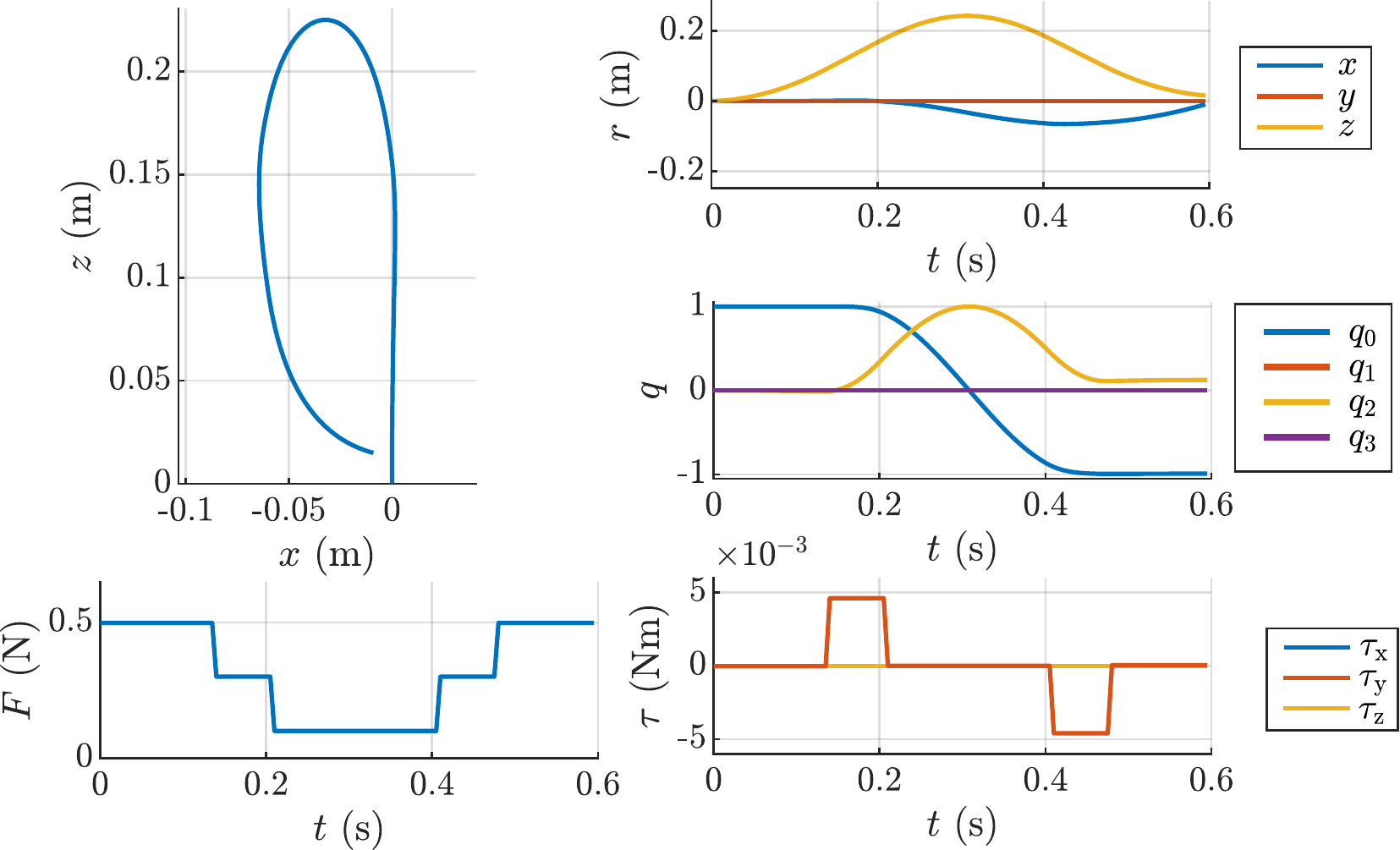}
\caption{Backflipping in simulation by feedforward control. The position ($r$), orientation ($q$), collective thrust ($F$), and input torque ($\tau$) are displayed.
}\label{fig:opensimu} 
\vspace{-4mm}
\end{figure}

The simulations have been executed using MuJoCo, a high-fidelity physics engine\footnote{\url{https://mujoco.org/}}. 
The implementation of GP regression is based on GPyTorch \cite{gardner2018gpytorch}, while the Bayesian Optimization is based on \cite{bayesopt}. All of the simulation code used in this work is available at our GitHub\footnote{\url{https://github.com/AIMotionLab-SZTAKI/crazyflie_backflipping}}, and a video illustrating the simulation results can be found at \url{https://youtu.be/Ed9jYlZr95c}. In this {and} the upcoming sections, we display obtained results with the $z$ axis pointing upwards (in contrast to the NED convention discussed in Section \ref{sec:model}), because the backflip maneuver is more illustrative this way.

\subsection{Bayesian Optimized Feedforward Control}\label{sec:opensimu}
\changed{The motion primitive parameters of the backflip maneuver have been calculated as the solution of Optimization~\eqref{eq:optim}, using 250 random initial function evaluations and 1000 iterations. The result of the numerical optimization is
\begin{align}\label{eq:optparam}
    \begin{split}
        p^*&=\begin{bmatrix}
U_1^* & t_1^* & t_3^* & U_5^*& t_5^*
\end{bmatrix} ^\top \\ & =  \begin{bmatrix}
17.8 & 0.14 & 0.2 & 17.8 & 0.12 
\end{bmatrix}^\top,
    \end{split}
\end{align}
where the unit of the collective accelerations $U_i^*$ are m/s$^2$, and the time is in seconds. 
Simulation results are displayed in Fig.~\ref{fig:opensimu}, using the optimized parameter vector given by \eqref{eq:optparam}. 
On the left plot, the position of the quadcopter during the flip is shown. At the end of the optimal maneuver, the position is $r=[-0.009,0,0.015]$~m with $[\phi,\theta,\psi]=[0,0.24,0]$~rad orientation, thus the elements of the final state error $e$ are uniformly small. 
On the right, the trajectory of the position vector, the orientation in quaternion representation, the collective thrust, and the input torques are shown, where the five phases of the maneuver defined in Section~\ref{sec:flip} can be clearly identified. The figure shows that almost near-maximal and near-minimal collective thrust and torque commands are required to perform the maneuver successfully. {Switching between these extreme values introduces} discontinuities of the control input {where} the unmodeled transient behaviour of the actuator dynamics can be significant. {The latter can influence the performance of the control strategy}. A possible solution would be to use a different parametrization of the control inputs instead of bang-bang control (e.g. spline parameters), however, such changes would make the feedforward design much more complex. 
}

\begin{figure}
  \centering
  \includegraphics[scale=.48]{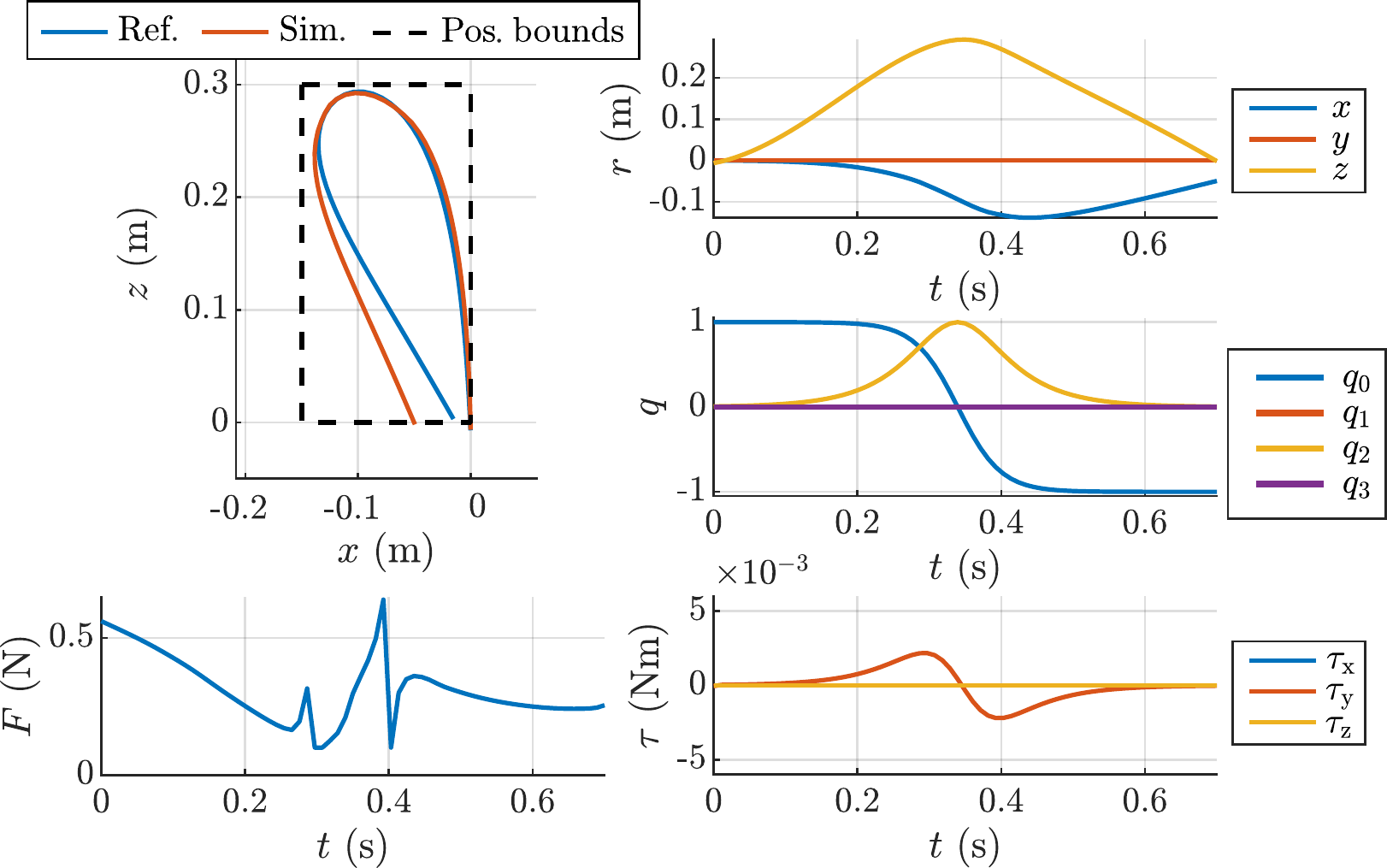}
  \caption{Backflipping in simulation by geometric control. The position, orientation, and control inputs are displayed.}\label{fig:geomsimu} \vspace{-2mm}
  \end{figure}
  
 \begin{figure}
  \centering
  \includegraphics[scale=.48]{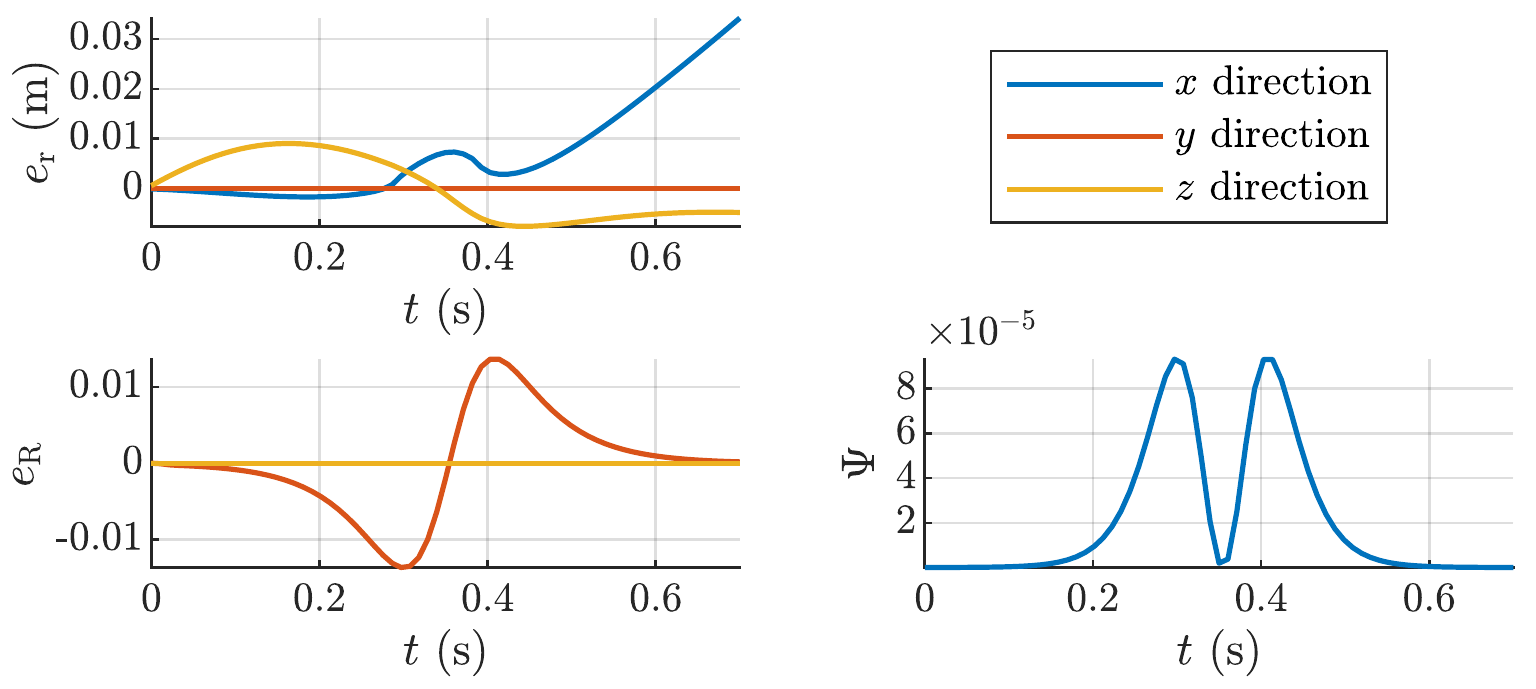}
  \caption{The trajectory of the error terms $e_\mathrm{r}, e_\mathrm{R}$ and the attitude error function $\Psi$ in simulation by geometric control.}\label{fig:geomsimuerror} \vspace{-4mm}
  \end{figure}

\subsection{Geometric Control and Trajectory Planning}
The second approach to perform a flip maneuver is trajectory planning and reference tracking with geometric control. Based on the results of the flip with feedforward control, the parameters of the reference pitch trajectory are chosen to $\nu_\mathrm{m}=35$~1/s, $t_\mathrm{m}=0.7$~s as illustrated in Fig.~\ref{fig:attref}. The quadratic optimization in \eqref{eq:quadprog} is solved under the following contraints:
\begin{align*}
    &\mathcal{X}:\;\{x_-, x_+, z_-, z_+ \} =  \{ -0.15, 0, 0, 0.3 \}\;\mathrm{m},\\
    &\mathcal{U}:\;F \in [0,0.64]\;\mathrm{N},
\end{align*}
with sampling time $T_\mathrm{s} = 2$ ms. Based on \cite{Forster}, the maximal collective thrust limit {is chosen to} $F_\mathrm{max}=0.64$ N {together with} position~bounds such that the trajectory is feasible and the quadcopter {exploits} the available flying space while avoiding collision with walls and obstacles. 
\changed{The quadratic optimization in \eqref{eq:quadprog} is solved off-line before starting the maneuver, on a desktop PC with Intel Core i9 processor and 16 GB of RAM. The computation time of the trajectory is 0.11 s in average, using Matlab with Mosek\footnote{\url{https://www.mosek.com/}}.}


   \begin{figure}
  \centering
  \includegraphics[scale=.48]{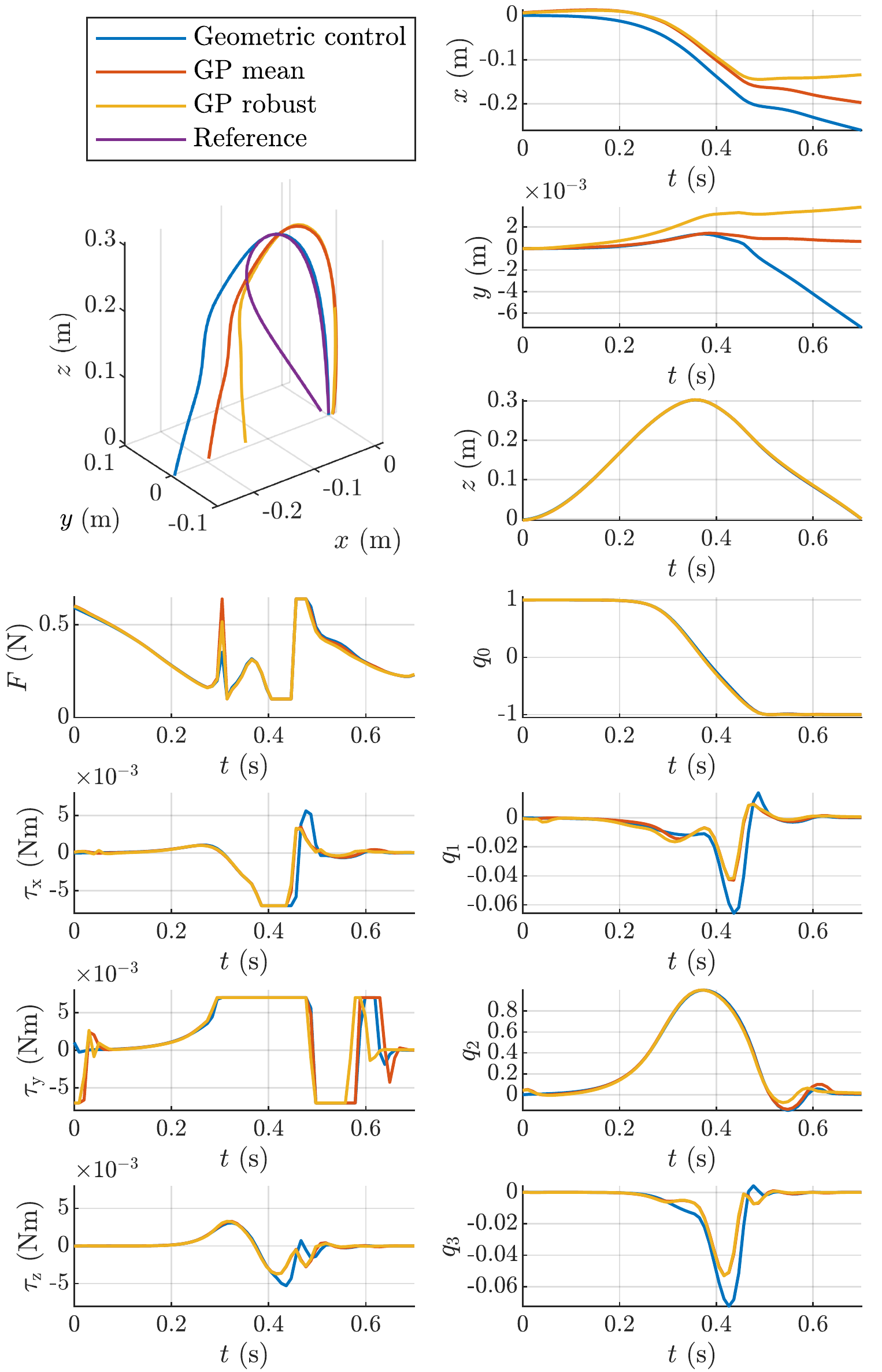}
  \caption{Backflipping in simulation using nominal, adaptive, and robust geometric control, with additional uncertainty. The position, orientation, and control inputs are displayed.}\label{fig:gp_simu} \vspace{-4mm}
  \end{figure}
  
     \begin{figure}
  \centering
  \includegraphics[scale=.48]{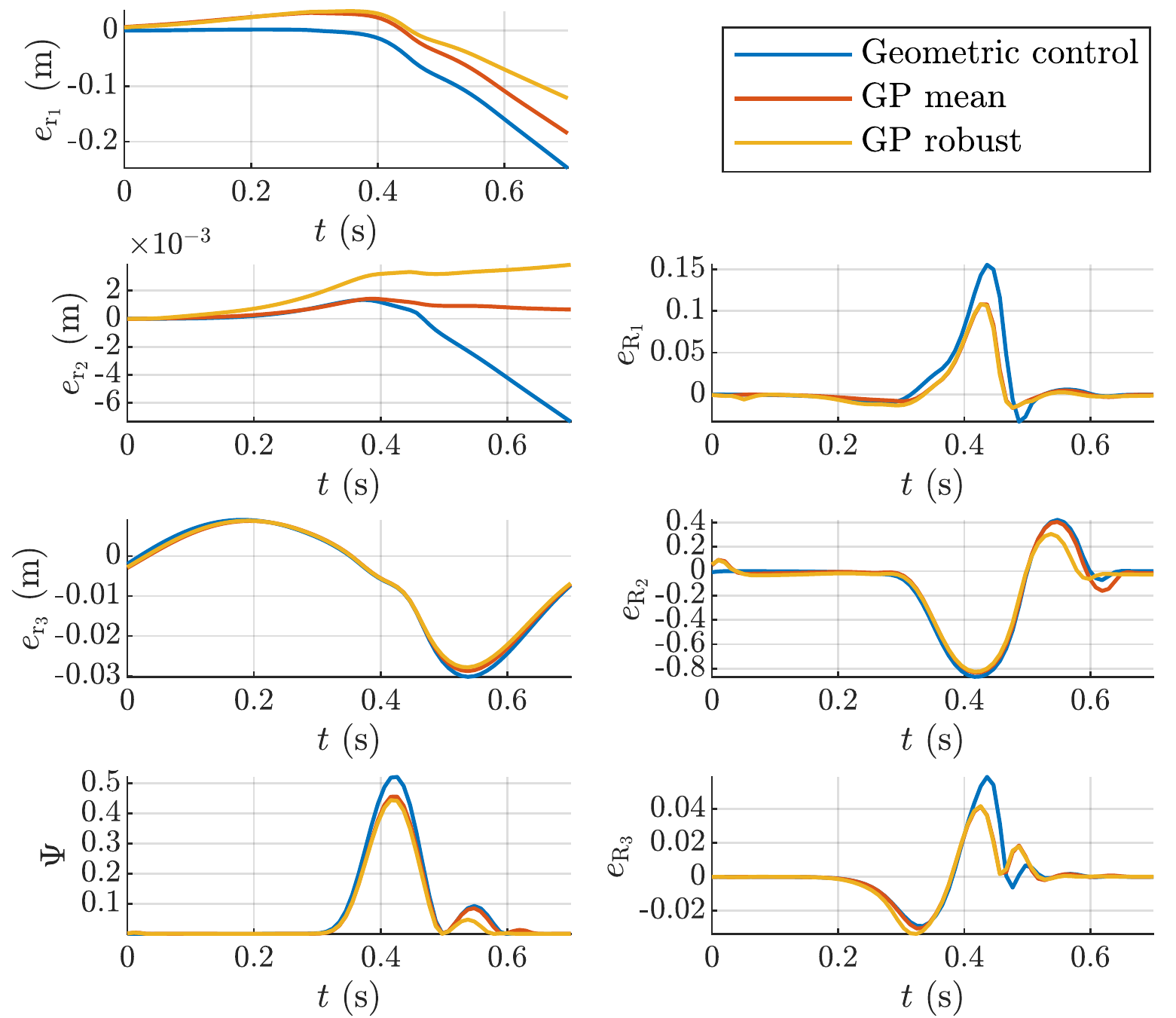}
  \caption{Comparison of the error terms $e_\mathrm{r}, e_\mathrm{R}$ and the attitude error function $\Psi$ in simulation with additional uncertainty.}\label{fig:geomrobsimuerror} \vspace{-2mm}
  \end{figure}

The gains of the geometric controller have been determined based on the stability conditions detailed in \cite{lelemc2010}, resulting in
\begin{align}\label{eq:gains}
  k_\mathrm{r} = 4.5,\; k_\mathrm{v} = 0.3,\;k_\mathrm{R} = 0.2,\;k_\omega = 0.002.   
\end{align}
\changed{The simulation results of the trajectory planning and reference tracking with nominal geometric control are displayed in Figs.~\ref{fig:geomsimu} and \ref{fig:geomsimuerror}. 
The trajectory of the control torque is smooth compared to the commands given by the feedforward controller, resulting in {less possible sensitivity w.r.t.} unmodeled actuator dynamics. Moreover, Fig.~\ref{fig:geomsimuerror} shows that {highly} accurate tracking of the attitude reference can be achieved ($\Psi$ {has the} order of magnitude of $10^{-5}$), {which also leads} to small position tracking errors.}

Next, we evaluate the performance of the proposed robust adaptive geometric controller for backflipping. We assume that the most significant modelling errors compared to the real drone arise in the rotational dynamics, due to the inaccuracy of the inertia matrix and center of gravity, and other aerodynamic effects. In simulation, we apply an external disturbance characterized by
\begin{align}
  \Delta_\mathrm{R} = \begin{bmatrix}
    -0.007 & -0.007 & 0
  \end{bmatrix}^{\!\top}\!\!\cdot \sin\! \left(\frac{\phi}{2} + \frac{\theta}{2}\right)\;\mathrm{Nm}.
\end{align}
We use the controller gains given by \eqref{eq:gains}, and choose the following constant values based on the stability conditions detailed in \cite{Lee2013NonlinearRT}: $\tau = 3, c_1 = 1, c_2 = 0.1, \epsilon_r = \epsilon_R = 0.0004$.

In the example of the backflip maneuver, we use only two scalar adaptive terms: the roll and pitch term of $\eta_\mathrm{R}$, namely $\eta_\mathrm{R, 1}$ and $\eta_\mathrm{R, 2}$, because most of the uncertainties arise in the roll and pitch motions. The adaptive terms are represented by two independent GPs with the following four-dimensional input: the $x$ and $y$ element of the attitude quaternions ($q_1, q_2$) and the angular velocity elements $\omega_\mathrm{x}$, $\omega_\mathrm{y}$. The adaptive GP terms introduced in Section~\ref{sec:geom} depend on the full state vector, however, in case of the backflip scenario, only these four inputs are relevant, and by reducing the input dimension, the model complexity is decreased radically without losing its expressiveness. For the GPs, we use zero mean and squared exponential covariance, given by \eqref{eq:sekernel}. The signal variance and lengthscale hyper parameters are trained using maximum likelihood estimation on 125 training points. Due to the relatively small input dimension and number of training points, the training and evaluation of the GPs are fast and efficient. 

\changed{The simulations by robust adaptive geometric control are shown in Figs. \ref{fig:gp_simu} and \ref{fig:geomrobsimuerror}. We compare the results using nominal geometric control (without adaptive and robust terms), geometric control with GP mean (without robust terms), and robust adaptive geometric control given by \eqref{eq:geomlaw}. 
Our results show that the attitude error of the nominal controller during backflipping is reduced significantly by using the proposed solutions, especially in terms of $e_{\mathrm{R_1}}$, $e_{\mathrm{R_3}}$. The position error is most significant in the $x$ direction ($e_{\mathrm{r_1}}$ is an order of magnitude larger than $e_{\mathrm{r_2}}, e_{\mathrm{r_3}}$), where the adaptive and robust controllers are capable to radically improve the tracking performance.}

 

%% file: sec7-experiments.tex
\section{Experimental Study}\label{sec:exp}
\subsection{Experimental Setup}
\begin{figure}
    \centering
    \includegraphics[width=.45\textwidth]{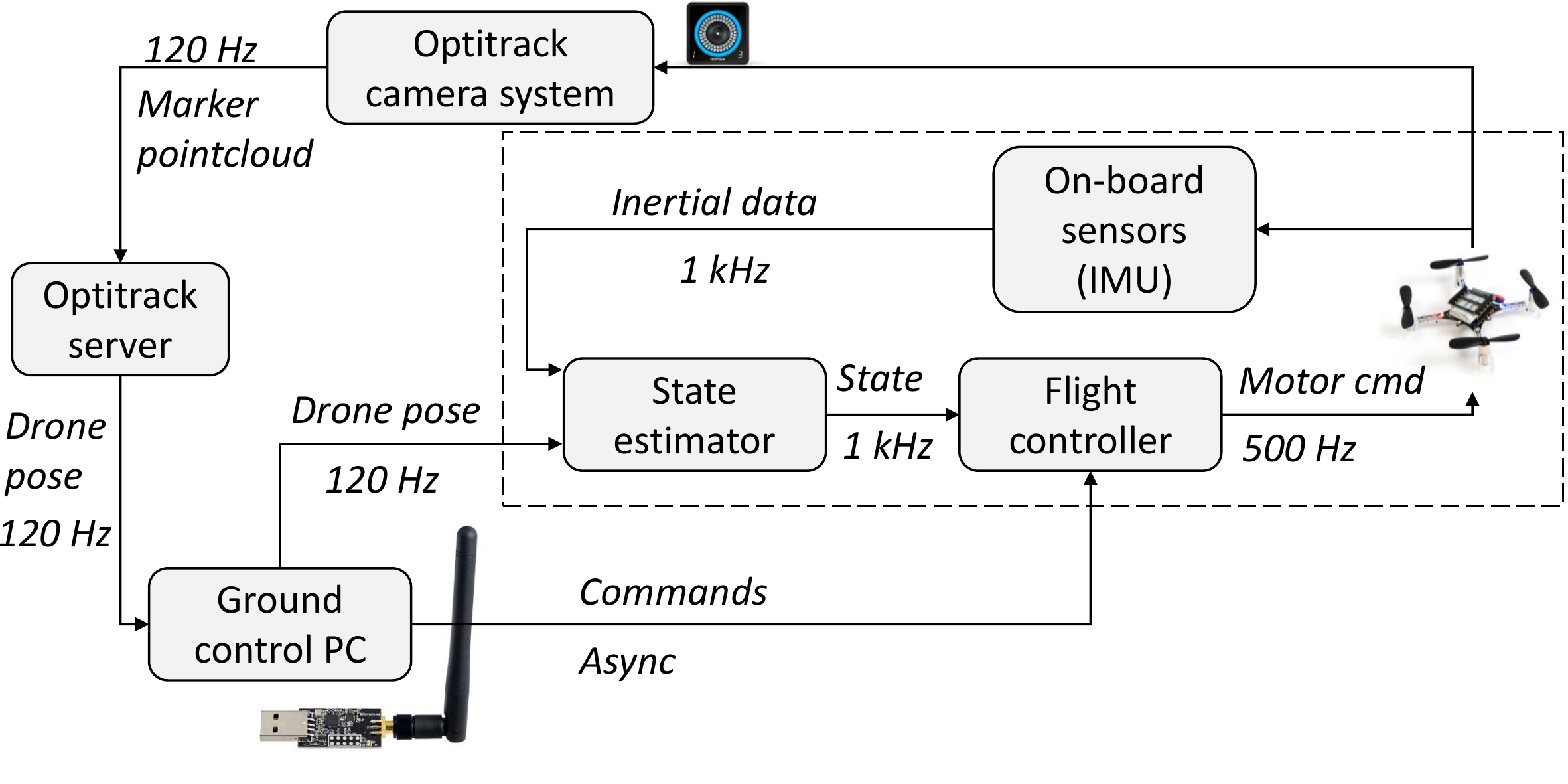}
    \caption[Block diagram of the experimental setup]{Block diagram of the experimental setup: indoor quadcopter navigation with internal and external localization.} \vspace{-4mm}
    \label{fig:sys}
    \end{figure}

The real experiments are performed with a Bitcraze Crazyflie 2.1  drone. Optitrack motion capture system\footnote{\url{https://optitrack.com}} is used to provide high precision position and orientation information. The drone and the positioning system is interconnected via a ground control PC, which runs the high-level experiment management and {executes} data-logging as well. The block diagram presenting the interconnection of the components is shown in Fig.~\ref{fig:sys}. The quadrotor is equipped with an IMU containing a 3D accelerometer, gyroscope, magnetometer and barometer, and it has two microcontrollers: a STM32F405 for running the flight controller, and a nRF51822 for radio communication and power management. Additionally, we use 
an expansion deck for high speed logging of measurement data to a micro SD-card. 
The quadcopter runs the original Bitcraze firmware augmented with our proposed control algorithms, while on the server, the Crazyswarm software platform is used to ease the implementation and configuration of high-level control components \cite{crazyswarm}. 

\subsection{Optimized Feedforward Control}

Firstly, we evaluate the results of performing the backflip with optimization-based feedforward control. For this experiment, the optimized parameter set given by \eqref{eq:optparam} is used. 
\begin{figure}
    \centering
    \includegraphics[scale=.45]{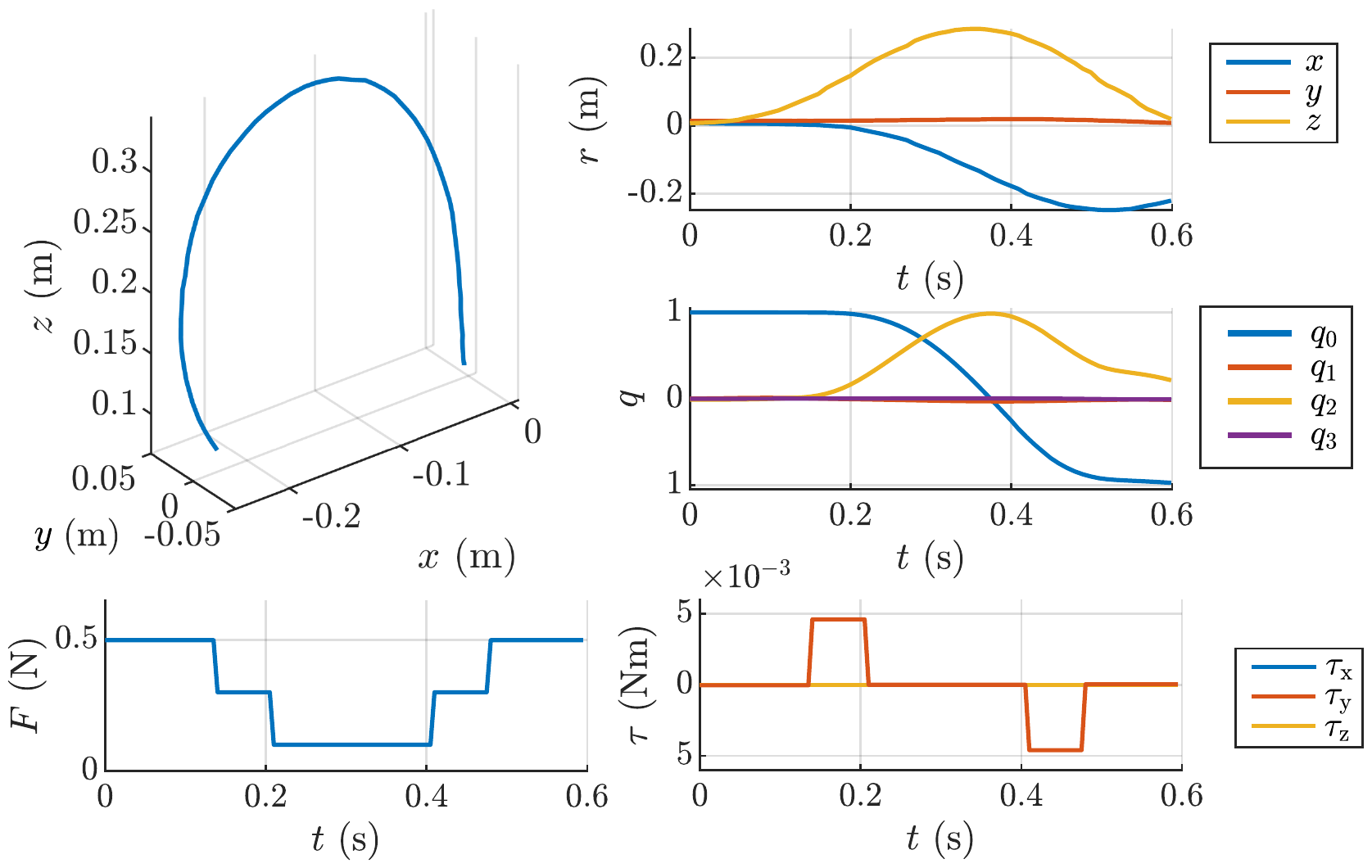}
    \caption{Backflipping measurement results with feedforward control using the optimized parameter vector given by \eqref{eq:optparam}. The position, orientation, and control inputs are displayed.}\label{fig:openmeas} \vspace{-2mm}
\end{figure}

\begin{figure}
\centering
\includegraphics[scale=.45]{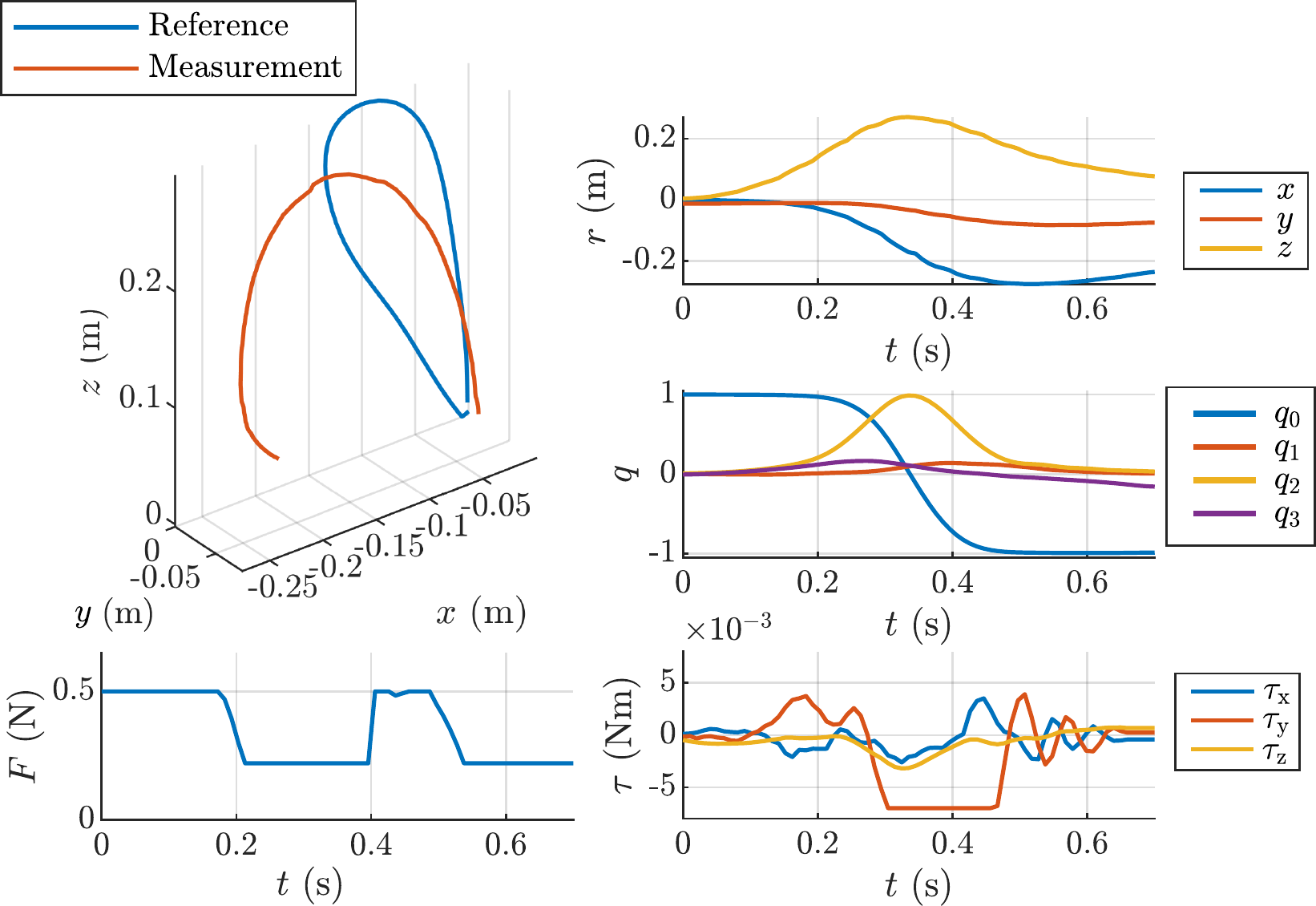}
\caption{Backflipping measurement results with nominal geometric control. The position, orientation, and control inputs are displayed.}\label{fig:geommeas} \vspace{-2mm}
\end{figure}    

\begin{figure}[!t]
\centering
\includegraphics[scale=.48]{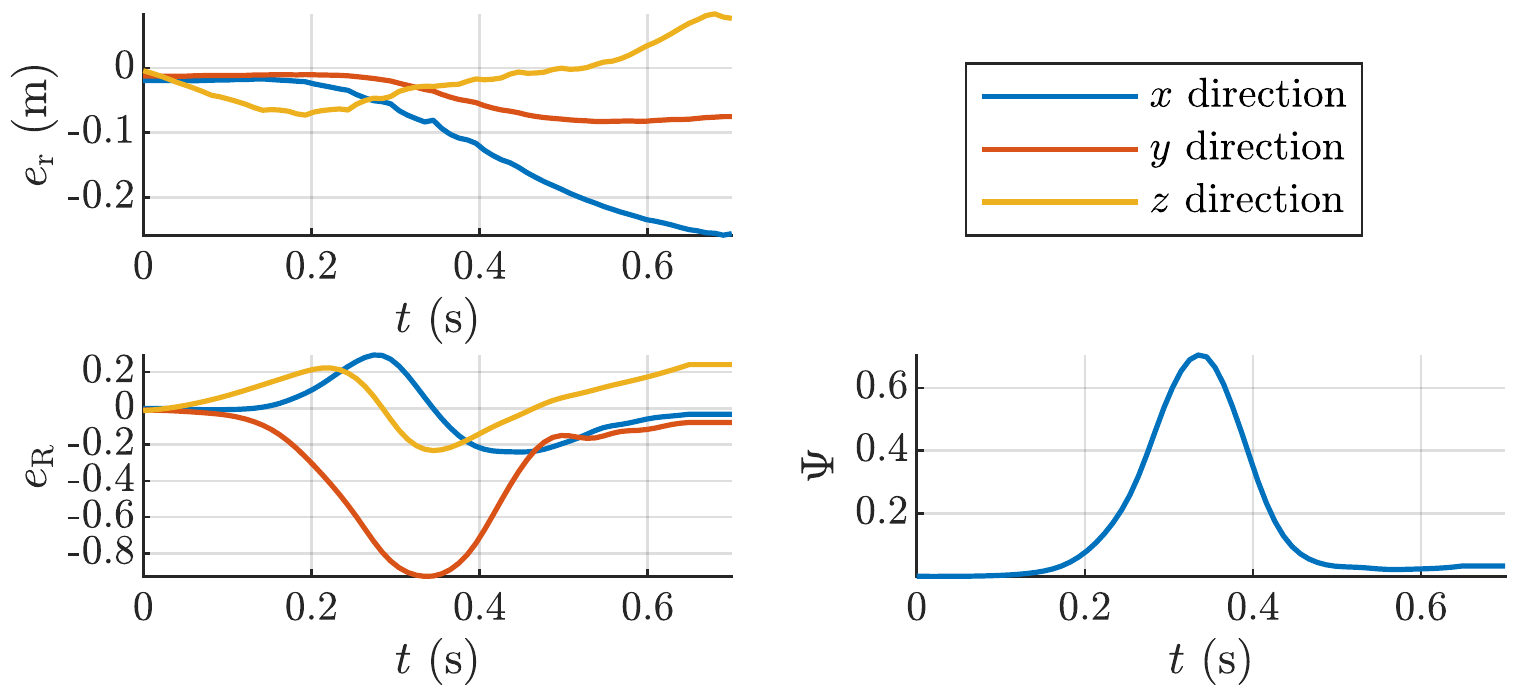}
\caption{The error terms $e_\mathrm{r}, e_\mathrm{R}$ and the attitude error function $\Psi$ in real experiments by geometric control.}\label{fig:geommeaserror} \vspace{-4mm}
\end{figure}   

\changed{The measurement results are displayed in Fig. \ref{fig:openmeas}, showing 
that the flip is executed with $[x\ y\ z] = [-0.22\ 0.008\ 0.01]$~m final position error and $[\phi\ \theta\ \psi]=[0.032\ 0.41\ 0.021]$~rad final error in Euler angles. Compared to the simulation displayed in Fig.~\ref{fig:opensimu}, the maximal displacement from the origin {in direction $x$} is around four times larger, {while} in direction {$z$ is} around 23\% larger. The difference is due to the uncertainties of the simulation model, however, the backflip is still performed successfully and the quadcopter is near the initial configuration at the end of the maneuver. It is important to note that the feedforward approach is sensitive to uncertainties in the dynamics and initial conditions. For example, if the flip maneuver begins when the orientation of the quadcopter is not horizontal, the stability can be lost at the recovery phase.}

\begin{figure}[!t]
    \centering
    \includegraphics[scale=.48]{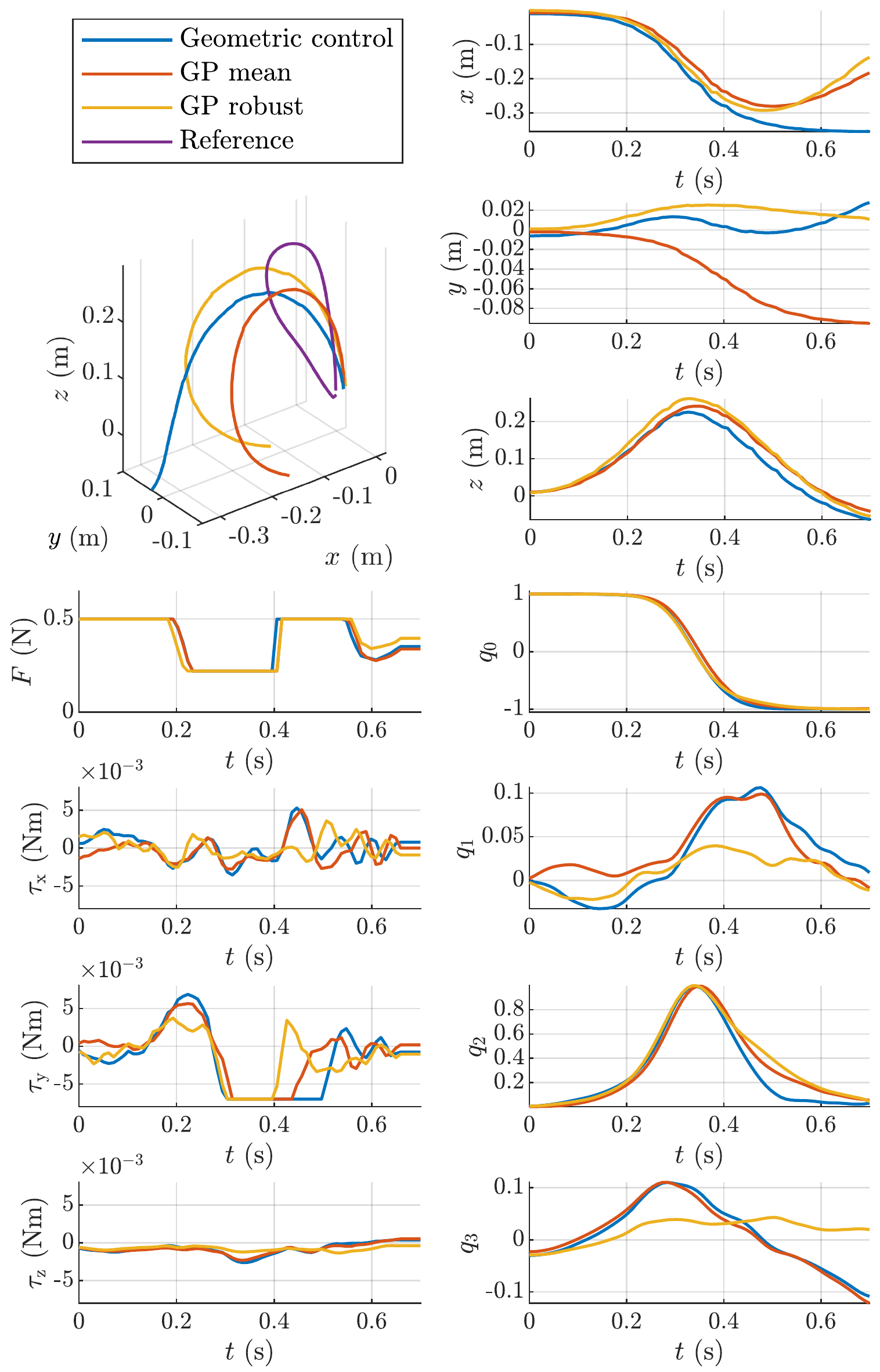}
    \caption{Measurement results with a bolt attached to the quadcopter, using nominal, adaptive and robust geometric control. The position, orientation, and control inputs are displayed.}
    \label{fig:gp_meas}\vspace{-3mm}
    \end{figure}
    
\begin{figure}[!t]
    \centering
    \includegraphics[scale=.48]{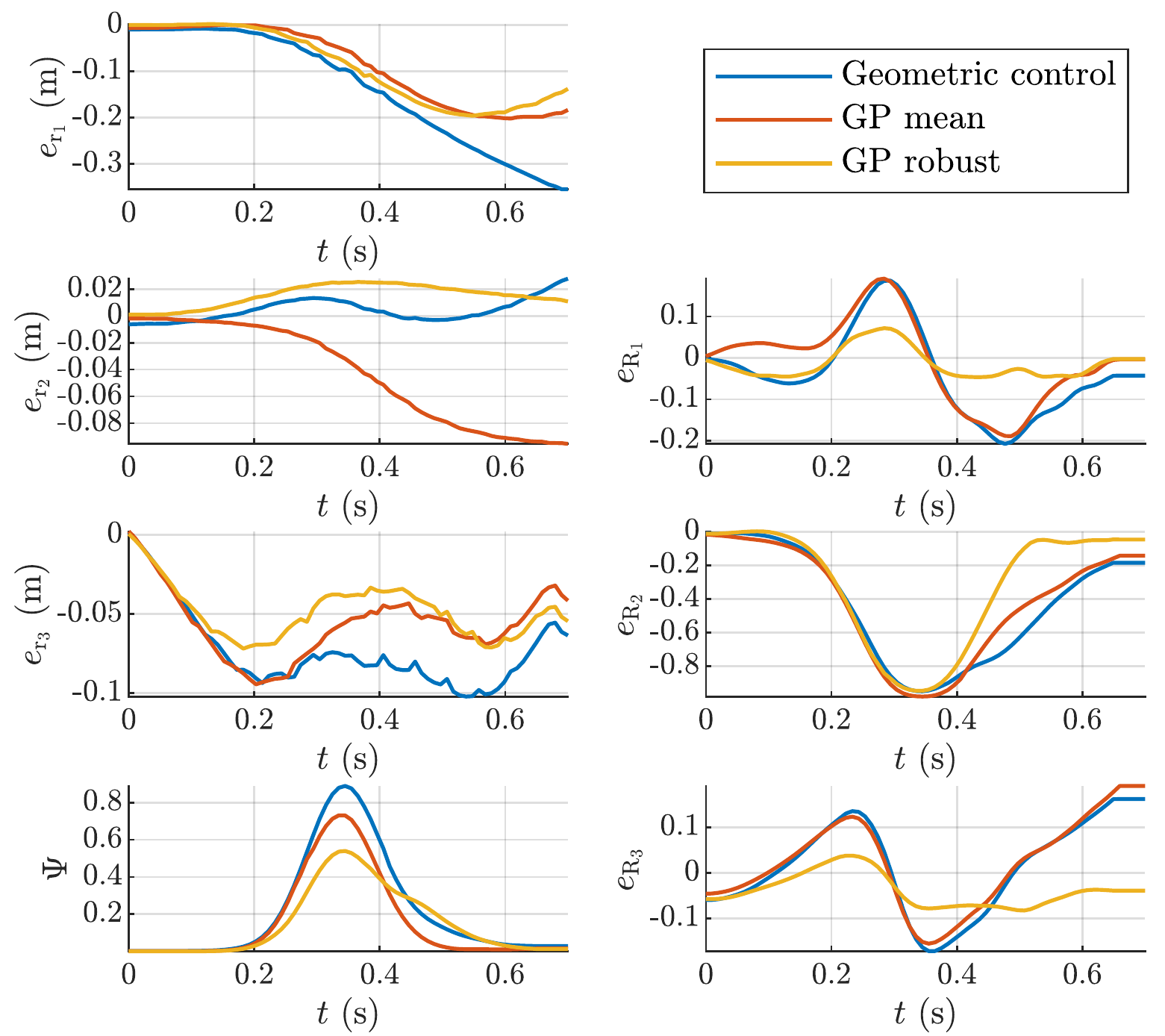}
    \caption{The trajectory of the error terms $e_\mathrm{r}, e_\mathrm{R}$ and the attitude error function $\Psi$ in real experiments with a bolt attached to the quadcopter.}
    \label{fig:gp_meas_error}\vspace{-5mm}
    \end{figure}

\subsection{Trajectory Planning and Geometric Control}

\changed{The experimental results of backflipping with the nominal geometric control are displayed in Figs.~\ref{fig:geommeas} and \ref{fig:geommeaserror}. The most important part of reference tracking is the attitude error function $\Psi$ and error vector $e_\mathrm{R}$, because a fast, stable and accurate attitude tracking is required to perform the flip maneuver and recover successfully. As it is shown on the left plot of the measurement results, the attitude error $e_\mathrm{R}$ is small in all directions and the controller remains stable. Although the position error $e_\mathrm{r}$ is larger (especially in the $x$ direction), the stability of the controller guarantees that the quadcopter gets back to the initial position after the backflip maneuver.}
In spite of the imperfect position tracking, the geometric controller is able to perform the backflip maneuver exactly the same way ten out of ten times, which indicates that even the nominal geometric control algorithm is significantly more robust than the feedforward method. 

\begin{figure}
\centering
\setlength{\fboxsep}{0pt}%
\setlength{\fboxrule}{0.5pt}%
\fbox{\includegraphics[width=.35\linewidth]{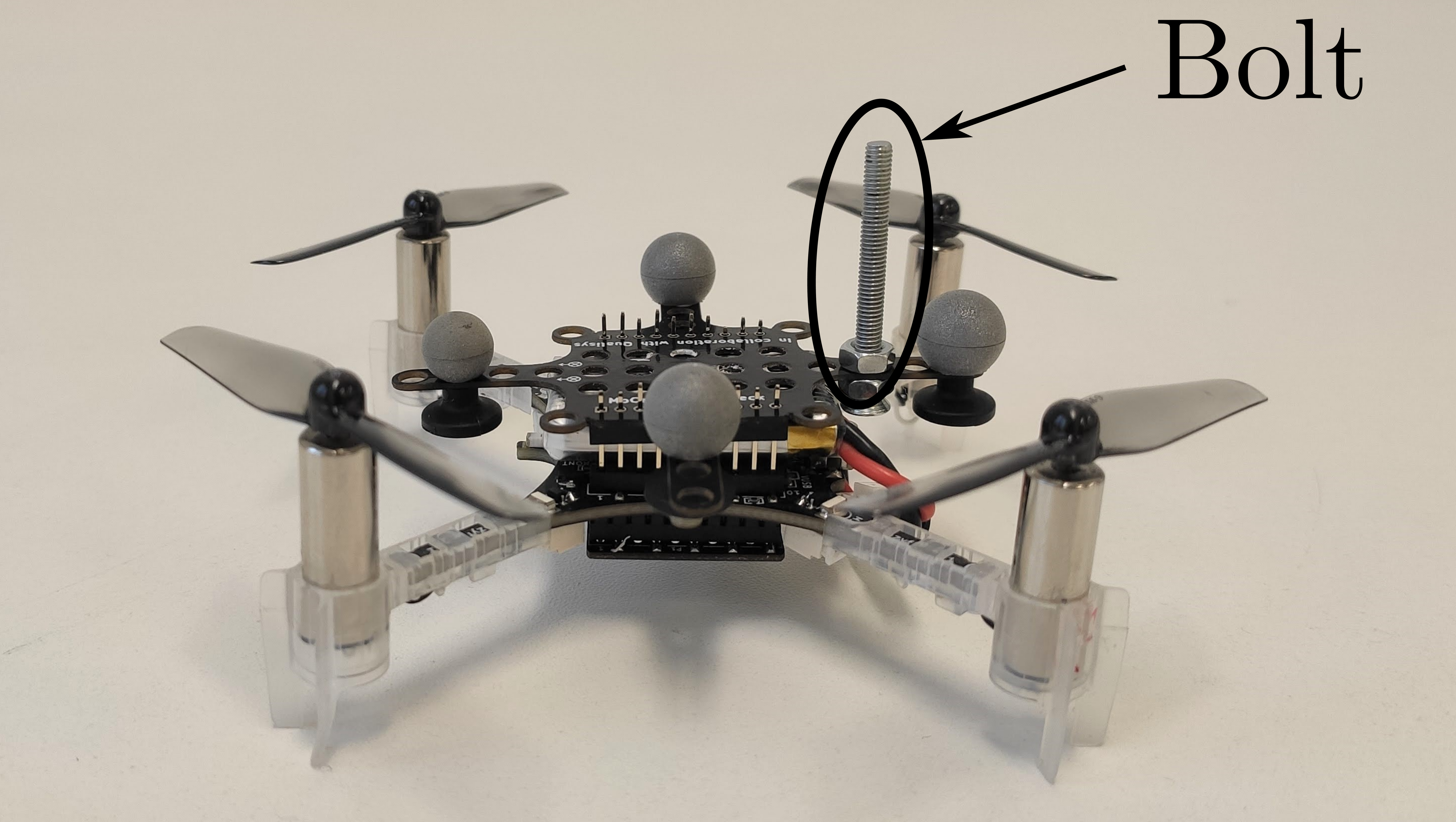}}
\caption{Crazyflie 2.1 quadcopter with reflective markers and a steel bolt used {to introduce significant additional dynamics}.}
\label{fig:bolt} \vspace{-1mm}
\end{figure}

\begin{figure}
    \setlength{\fboxsep}{0pt}%
    \setlength{\fboxrule}{0.5pt}
    \centering
    \begin{subfigure}{.45\linewidth}
      \centering
      \fbox{\includegraphics[height=3cm]{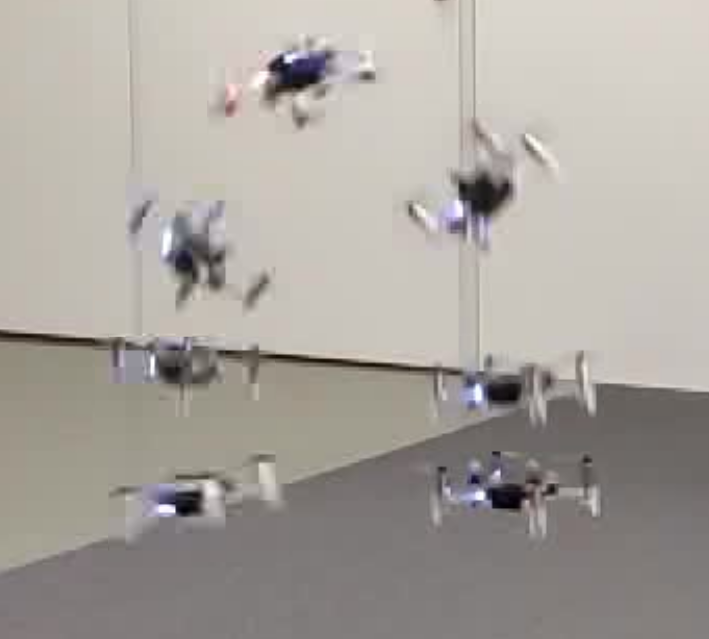}}
      \caption{Optimization-based feedforward control.}
      \label{fig:sub1}
    \end{subfigure}%
    \hspace{0.3cm}
    \begin{subfigure}{.45\linewidth}
      \centering
      \fbox{\includegraphics[height=3cm]{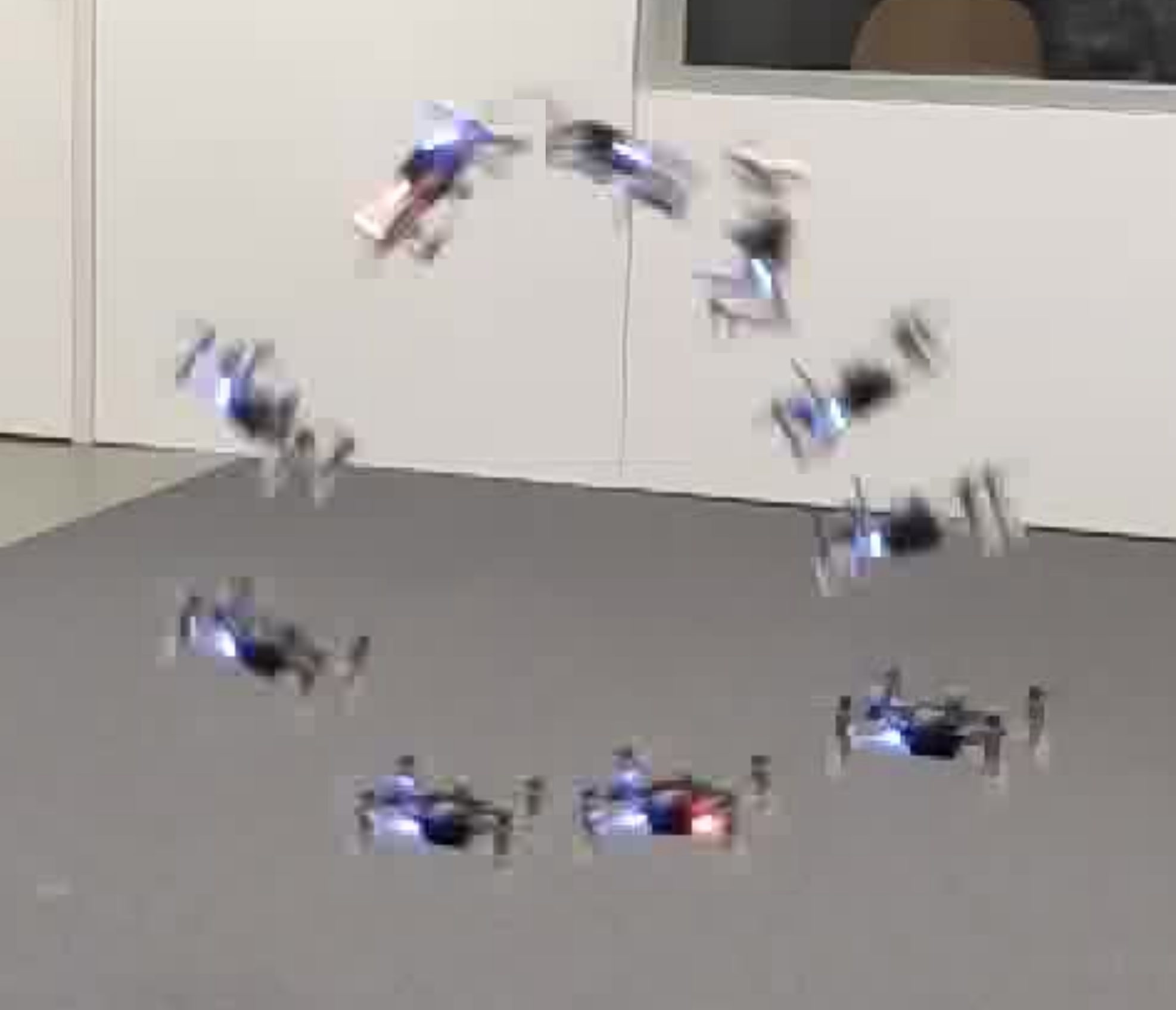}}
      \caption{Robust geometric feedback control.}
      \label{fig:sub2}
    \end{subfigure}
    \caption{Composite images of the experiments using a Bitcraze Crazyflie 2.1 quadcopter.} \vspace{-4mm}
    \label{fig:test}
    \end{figure}

\changed{Next, we evaluate the performance of the proposed robust adaptive geometric controller with model uncertainty added to the quadcopter dynamics. We have found that a steel bolt attached to the drone (as illustrated in Fig.~\ref{fig:bolt}) has significant influence on the attitude dynamics, however, the vehicle is still able to perform the backflip maneuver. Using the modified configuration, we collect data points by performing agile maneuvers with the nominal geometric controller and fit a Gaussian Process on the measurement data. During the flights, the GP has to be evaluated at 500 Hz, which is not possible due to the limited computational capacity of the on-board microcontroller unit. Therefore, we generate a lookup table by evaluating the trained GP on a grid of the input variables with 2025 grid points off-line and upload it to the on-board {microcontroller} of the quadcopter. 
Another viable alternative would be the use of sparse GP methods, {e.g.,} \cite{RW2006,unifying_sparse}), which are possible to evaluate real-time, especially on larger quadcopter platforms with more powerful computational unit (e.g. NVIDIA Jetson \cite{Foehn2022}).}

\changed{The measurement results are displayed in Figs.~\ref{fig:gp_meas} and \ref{fig:gp_meas_error}. Our results show that using the proposed adaptive and robust controllers, the attitude tracking error is even more reduced in real flights than it is in simulation. Moreover, the position tracking performance is also enhanced by the robust approach, especially in the $x$ and $z$ directions. 
}


%% file: sec8-summary.tex
\section{Conclusion}\label{sec:conc}
\changed{In this paper, a Bayesian optimization based feedforward control and {a} robust geometric reference tracking control {approach} with optimization-based trajectory planning have been proposed for quadcopters to reliably perform the backflip maneuver in the presence of modeling uncertainties. In the former, relatively simple approach, Bayesian optimization can be used to fine tune the feedforward sequence, leading to efficient implementation both in simulation and real flights. The second method utilizes Gaussian Process based augmented motion models that are able to precisely approximate model uncertainties. Combined with geometric control, the resulting architecture provides robust stability and high control performance, outperforming the feedforward method both in terms of reliability and optimal tracking of the motion profile.}

\changed{In our future research, we intend to use learning methods to perform complex maneuvers with less expert knowledge and extend the capabilities of the miniature drones even more.}